\documentclass[12pt,preprint]{aastex}
\usepackage{graphicx}
\DeclareMathAlphabet{\mathpzc}{OT1}{pzc}{m}{it}

\slugcomment{}

\shorttitle{}
\shortauthors{}

\begin{document}

\newcommand{\hh}{H{\small II} }

\title{A New Catalog of H{\small II} Regions in M31}
\author{M. Azimlu, R. Marciniak, \& P. Barmby}
\affil{Department of Physics \& Astronomy\\
University of Western Ontario\\
1151 Richmond St., London, ON N6A 3K7 Canada}

\begin{abstract}
We present a new catalog of \hh regions in M31. The full disk of the galaxy ($\sim$24 kpc from the galaxy center) is covered in a 2.2~deg$^{2}$   mosaic of  10 fields observed 
with the Mosaic Camera on  the Mayall 4 m telescope as part of the Local Group Galaxies survey. 
We used HIIphot, a code for automated photometry of \hh regions, to identify the regions and measure their fluxes and sizes. 
A $10\sigma$ detection level was used to exclude diffuse gas fluctuations and star residuals after continuum subtraction. That selection limit may result in missing  some faint \hh regions, but our catalog of 3691  \hh regions is still complete to a luminosity of L$_{H\alpha} =10^{34}$~erg~s$^{-1}$.
This is five times fainter than the only previous CCD-based study which contained  967 objects in the NE half of M31. We determined the H$\alpha$ luminosity function (LF) by fitting a power law to luminosities larger than L$_{H\alpha}=10^{36.7}$ and determined a slope of 2.52$\pm$0.07.  
The in-arm and inter-arm LFs peak at different luminosities but they have similar bright-end slopes.
The inter-arm regions are less populated (40\% of total detected regions) and constitute only 14\%   of the total luminosity of L$_{H\alpha} = 5.6\times 10^{40} $~erg~s$^{-1}$ (after extinction correction and considering  65\% contribution from diffused ionized gas). 
A star formation rate of 0.44~M$_\odot$~yr$^{-1}$ was estimated from the H$\alpha$ total luminosity; this value is consistent with the determination from the {\em Spitzer}  8~$\mu$m image.
We removed all known and potential planetary nebulae, yet we found a double peaked luminosity function. The inter-arm older population suggests a starburst between 15 and 20 million years ago. 
This result is in agreement with UV studies of the star formation history in M31 which found a star formation rate decrease in the recent past.
 We found a  fair spatial correlation between the \hh regions and stellar clusters in selected star forming regions.  Most of the matched regions lie within the arm regions. 
 
\end{abstract}

\section{Introduction} 
\label{intro}

Dense clumps in giant molecular clouds  are the birthplaces of stars in galaxies. Most of the newborn stars are embedded within  dense cores and obscured by dust. Only massive stars  can  heat the gas and dust in their environs  by emitting ultraviolet (UV) photons  and ionizing the surrounding  hydrogen, producing  \hh regions. The ionized gas radiates mostly  in the H$\alpha$ line, therefore photometry and spectroscopy of H$\alpha$ emission is one of the main probes of local and global star formation in nearby galaxies. \hh regions are one of the best-known massive  star formation tracers in other galaxies \citep[e.g.][]{kennicutt08, thilk2000, lawton10}. They might be excited by only one  star or multiple young massive stars or even clusters. Properties of \hh regions such as shape, size, distribution and abundances vary with the physical characteristics of  the ionizing stars, galaxy type, and environmental conditions such as  gas density and metallicity \citep{hodge83, kennicutt84,  elmegreen2000, thilk02, esteban09}. 

\hh regions are  observed to have a wide range of size and luminosities. For example 30 Doradus  (size $\sim$300 pc) in the Large Magellanic Cloud (LMC) is the largest known \hh region in the Local Group and is excited by a cluster of massive stars with a total stellar mass of $0.35-1\times10^5$~M$_\odot$ \citep{Campbell2010}. In comparison, a typical Galactic \hh region is excited by an individual B star and can be smaller than a parsec. 

The characteristics of \hh regions have been studied in many extragalactic H$\alpha$ surveys [e.g. \citet{knapen98} and references therein; \citet{petit98, thilk02,  kennicutt08}]. 
The studies by \citet{kennicutt88} and \citet[][KEH89 hereafter]{kennicutt89}  of the population of \hh regions in nearby galaxies confirmed that  \hh region population  properties are strongly dependent on the host galaxy Hubble type. The massive star formation rate and the frequency of giant \hh regions are noticeably higher in late type galaxies. 

The  star formation properties of galaxies  can be determined through study of their numerous \hh  regions \citep[][WB92 hereafter]{walterbos92}.  From the H$\alpha$ flux one can determine the number of Lyman continuum photons \citep{spitzer78}, which in turn is used to determine the mass of the ionizing stellar population, providing an upper limit to the initial mass function (IMF) for massive stars \citep{oey03}.  Since the amount of H$\alpha$ luminosity depends on the ionizing flux of the young O and B stars, we expect that the \hh regions' luminosity function (LF) should also trace  the distribution of ionizing stellar masses \citep{oey03}.   The LF can be well fitted for all galaxies by a power law function but the slope varies with galaxy Hubble type (KEH89). The  \hh regions are mostly located  within the arms of spiral galaxies but the LF for in-arm and inter-arm regions might be different \citep[e.g.][]{thilk2000}. A Monte Carlo simulation    to study the evolution of \hh  LFs \citep{oey98} suggested that arm populations represent the current active star forming regions while inter-arm regions are aged populations.
However this method of studying the stellar populations only gives accurate measurements for luminous massive stars in nearby and face-on galaxies.
The size distribution of \hh regions in a galaxy can also be compared between galaxies as a structural  property \citep{bergh}.

M31 is the nearest large galaxy to our own and can be studied in great detail. \citet{baade64} made the first catalog of the position of emission nebulae in M31. The study was followed by that of \citet{arp73}, who measured the size of the largest extended regions in M31 and M33. 
The most recent complete catalog of \hh regions in M31 was produced by \citet{pellet78}. That study had a spatial resolution of 4\arcsec\ ($\sim 15 $pc) and detected 981 \hh regions. WB92  made the first deep CCD images of the North-East half of the galaxy in the H${\alpha}$ and $[SII]$ emission lines. They imaged 19 fields, each covering a 6.6 arcmin$^2$ area (the largest possible field of view at the time), trying to cover the most active star forming regions in the North-East half of M31. They measured and calculated the positions, dimensions and fluxes of 967  \hh regions. 

The complicated morphology of M31 suggests a violent history. A star forming ring with radius of 10 kpc which is not centered at the galaxy nucleus \citep{gordon2006}, and a second inner dust ring  0.5 kpc from the center  \citep{block2006} suggest past collisions. Other features such as tidal streams \citep[e.g.][]{ibata01,fardal08} indicate active interactions between M31 and other Local Group members. 

We  study the properties of the spatially resolved \hh regions in M31 and try to connect these to present and past star formation processes in the galaxy. 
In this work we present a new catalog of \hh regions over the entire disk of M31, using higher resolution ($1\arcsec, \sim$ 3.8 pc) H$\alpha$ imaging  
\citep{massey06}.  We identified 3961  distinct regions to a limiting size of $4.08\arcsec$ (15.6 pc) and limiting detection  flux  of   $\sim 10^{-16}$~erg~cm$^{-2}$~s$^{-1}$. Throughout this work we assume a distance to M31 of 0.783 Mpc \citep{Distance}. 

Details of the data set, data reduction and data analysis methods including the HIIphot code  parameters setting are provided in \S\ref{data}. In \S\ref{catalog}, we introduce the characteristics of the new catalog of \hh regions in M31 and compare it with previous works. In \S\ref{discussion} we derive the luminosity function, total H$\alpha$ luminosity, size distribution, and match our catalog with young stellar clusters in M31. \S\ref{summary} contains a  summary of our results. 

\section{Data Analysis} 
\label{data}

The data used came from the Nearby Galaxies Survey of \citet{massey06} and were obtained from the NOAO science archive. The images were taken at the 4 m Mayall telescope with the Mosaic CCD Camera between August 2000 and September  2002 and comprise
H$\alpha$ and R band mosaics of ten overlapping fields across the disk of M31.  The new images permit photometry with negligible uncertainty to an H$\alpha$ magnitude of 20 and have delivered image quality  varying between 0.9\arcsec  and 1.4\arcsec\ (equivalent to 3.4--5.3 pc). Each field has an approximate angular size of 36\arcmin $\times 36$\arcmin\ observed as a set of 5 dithered exposures, while the entire survey covered 2.2 square degrees of M31.  The pixel scale is $0\farcs258$ pixel$^{-1}$,  with an average PSF FWHM of 1\arcsec\ ($\approx$4 pixels).  
The accuracy of the astrometric calibration is not discussed by \citet{massey06}, although the discussion in \citet{massey07} implies
that it is likely to be good to 0\farcs1. That paper also states that photometric calibration goal for the narrow-line imaging was a precision
of 5--10\%, although it is not entirely clear from the discussion whether that goal was reached.

\begin{figure}
	\centering
	\includegraphics[width = 14cm]{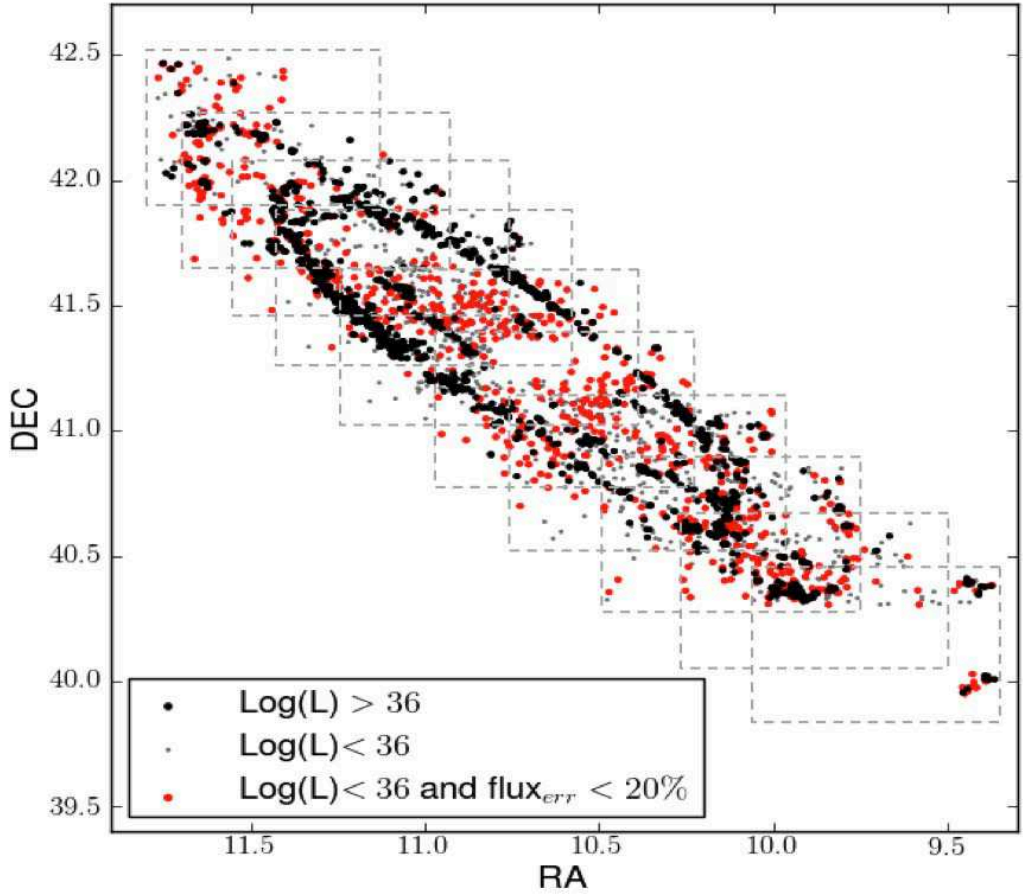}
	
\caption{ Coordinate plot of \hh regions in M31, with north up and east left.  Dashed rectangles  show the 10 fields observed by \citet{massey06} numbered from top to bottom.  The black dots indicate the regions with H$\alpha$ luminosities $\geq 10^{36} $ erg  s$^{-1}$. Grey dots are regions with luminosities   $ L_{H\alpha}< 10^{36} $ erg  s$^{-1}$ and red dots present the same regions with a final flux uncertainty  less than 20\%.  The central bulge is masked with a $10'\times10'$ box. The higher luminosity regions clearly trace the spiral arms while the lower luminosity regions also fill up the inter-arm spaces with a concentration toward the center. } \label{regions1}
\end{figure}

Figure \ref{regions1} shows the observed fields in dashed rectangles.  
A 10\arcmin $\times 10$\arcmin\ central region was partially saturated in the images and was
omitted from our survey.
The black dots show the detected \hh regions in our survey with  luminosities $\geq 10^{36} $ erg  s$^{-1}$. Red dots are regions fainter than this limit but with a flux uncertainty  smaller than 20\%. Grey dots present the remained faint regions.   The higher luminosity regions clearly trace the spiral arms while the lower luminosity regions  fill up the inter-arm spaces as well. 
We discuss the \hh regions' spatial distribution in more detail in \S\ref{clusters}.

Making continuum-subtracted images requires  removal of the stellar emission from the H$\alpha$ images using the R band image of the same field. To make the subtraction  we have to assume that all stars have the same H$\alpha$ fraction in their spectra.  This is not accurate as different spectral types have different absorption line depths for H$\alpha$, and what is being measured is the best fit  H$\alpha$  fraction so that most stars are properly removed.  Some residuals will be present due to variations in spectral shape and absorption line depths from star to star.  To perform the subtraction the amount of continuum emission present in each H$\alpha$ image must be determined. We used SExtractor to obtain the photometry of stars in  H$\alpha$ and R bands. A scaling factor 
between H$\alpha$ and R fluxes  was then obtained by comparing fluxes of the bright stars.  
All the images  had nearly a linear relation between the two bands with similar slopes (scaling factor) of  0.36 to 0.41;
Field 2 was an exception, with a scaling factor of 0.86. This field was reported by \citet{massey06} to have a flatness problem which resulted
in a slightly different color, so the different scaling factor is not unexpected.
The H$\alpha$ and R  images are aligned on the same coordinate grid, so for each field the continuum-subtracted H$\alpha$ image was constructed by 
subtracting the R-band image multiplied by the scaling factor from the H$\alpha$ image.
The scaling factors are reported in Table~\ref{scale} for the benefit of future users.

Calibration of the Nearby Galaxies Survey images is the final step before running H{\small II}phot.
H{\small II}phot requires that the continuum-subtracted image be in units of emission measure (EM). We convert 
from image units to EM in pc~cm$^{-6}$ using the image exposure time (300~s), the calibration factor given by
\citet{massey07} \citep[and the correction to its units given by][]{relano09}, the average pixel scale of 0\farcs258,
and the H$\alpha$-to-EM conversion ratio as used in the H{\small II}phot code of 
$2.0\times10^{-18}$~erg~cm$^{-2}$~s$^{-1}$~arcsec$^{-2}$~pc$^{-1}$~cm$^6$. 
Combining these factors gives an overall conversion factor from ADU to EM of 4.437.
We also need to account for the flux contribution to the H$\alpha$ filter from the [NII] 6583~\AA\ line. 
The [NII] percentage is given as the fraction of [NII] present in the filter band pass multiplied by the relative intensity of the [NII] line compared to H$\alpha$.  
\citet{james05} studied 334 nearby galaxies and found  that [NII] emission is not distributed exactly the same as the H$\alpha$ emission.  Therefore the 
I[NII]6583/I[H$\alpha$] ratio varies within a galaxy, particularly in nuclear regions and also with galaxy type.     
\citet{greenawalt97}, however, reported that in diffuse ionized gas (DIG) near \hh regions, the intensity ratio I[NII]6583/I[H$\alpha$] = 0.35 and we use this value for 
the [NII] contribution to the total detected emission.

\subsection{Running HIIPhot} 
\label{h2phot}

\hh regions come in a large range of sizes and shapes.  The characteristics of  \hh regions depend on the exciting stars, the ionizing photons they emit  and on the environment in which the stars are ionizing. Therefore \hh region photometry is more difficult than stellar photometry.
The physical extent of regions are not well defined, and borders between regions and emission from DIG are 
difficult to distinguish.  Different methods and algorithms have been used to identify the \hh regions, define the boundaries, subtract the background emission and measure the flux in various studies from hand works to detailed sensitive computer codes  \citep[e.g.][]{pellet78, knapen93,walterbos94,knapen98,rozas99, pleuss2000,scoville01,james04,gutier11}. 
H{\small II}phot  is an IDL code which was developed by \citet{thilk2000} for the sole purpose of performing accurate photometry of \hh regions in external galaxies. \citet{thilk2000}  applied this method to the spiral galaxy M51 and compared their resulting LF to that of \citet{rand92}, showing that H{\small II}phot can successfully and accurately reproduce and improve upon existing LFs.  

The detailed operation of H{\small II}phot  is described by \citet{thilk2000}; here we give a brief summary.
H{\small II}phot procedes in an iterative fashion. 

An appropriately-sized Gaussian fit is assigned to each local maximum of the data to remove source structure on smaller scales.  
Following this initial detection of sources, H{\small II}phot assigns {\em footprints} to these flux peaks, and considers as possible
detections only those Gaussian peaks are a specified multiple of the image noise.
This multiple is called ``S/N" in the H{\small II}phot input parameters, but this ``detection signal-to-noise" is not exactly equivalent to the
conventional definition as an object's flux divided by its flux uncertainty.
We determined the noise by integrating and averaging over the entire image, rather than a selected empty region, to make sure we did not 
underestimate the noise in crowded regions or bright regions adjacent to the bulge. 
 
After defining footprints and refining a list of regions, all pixels below 50\% of the footprintÕs median flux are rejected, creating seeds, which are a starting point for region growth.  Iterative growth proceeds in a manner where the region boundary only expands to adjacent pixels that are above a threshold level.  If the pixels are below the threshold they are ignored until a later iteration.  If less than 50\% of the boundary pixels are above the threshold, the region will not expand during the given iteration.  A region's growth ceases when its surface brightness profile slope flattens to below a user-specified threshold, or when 
it is completely surrounded by other region boundaries and cannot claim any more pixels.

Preparing the images carefully is critical in order for the code to produce reasonable results with a reasonable amount of computing time.
For example, in test runs with S/N=5, stellar residuals left over from the continuum subtraction process accounted for around 80$\%$ of total detections.  
We found that masking isolated stellar sources greatly reduced the run time. A source was determined to be
isolated if the flux in an annulus around it was comparable to the background flux in the image: masking only
isolated stars ensured that H{\small II} region emission fluxes and sizes were not affected.
The masked region around each star was a circle with size  
determined by the original continuum flux of the star, as brighter field stars leave larger residuals.  
After applying this correction  the number of false detections decreased to approximately 20$\%$ for S/N=5 and less than $1\%$ for S/N=10.  
We found that the original images required too much memory to be processed all at once; images for each field were cropped into 8 smaller segments with 250 pixel overlaps to make sure all large areas of H$\alpha$ emission were within a single segment. 
Duplicate objects detected in more than one segment were removed from the final catalog.

The results of the H{\small II}phot code are sensitive to many factors that determine both the specific results and the computing time needed to run the code.  
The 20 input parameters required for H{\small II}phot determine the range in H{\small II} region model sizes, the resulting fluxes, and the physical extent of region growth i.e. the definition of the boundary between region emission and surrounding DIG.  
The  distance to the galaxy is well established at 0.783 Mpc, as reported by \citet{Distance}.  The coordinates for the noise estimates were taken as the entire images.  This results in an overestimate for the noise due to inclusion of bright regions in the averaging.  However, the noise estimate is used for locating regions, and not in the flux determination, so the slight overestimate actually reduces the number of false detections due to isolated DIG. 
The last few input parameters describe the growth of regions, specifically the stopping point for growth in EM~pc$^{-1}$.  We adopt a value of 1.5 EM~pc$^{-1}$ for this terminal gradient, keeping consistency with the stopping point of \citet{thilk2000}, described as the boundary between the \hh region and the surrounding DIG. 

Different S/N and PSF detection limits were tested and the resulting luminosity and size distributions and individual luminosities were compared with WB92.  Two different sample fields, one at the edge of the galaxy (sub-field of F1) and one at the center (sub-field of F5)  were selected for the test runs. We found that increasing the S/N detection limit with the same PSF decreased  the number of detected faint regions but did not affect the LF slope,  because all the faint regions had luminosities smaller than the turnover peak used to determine the LF. On the other hand, decreasing  the PSF-FWHM 
with the same S/N resulted in fewer bright regions (they were broken into smaller sub-regions) and resulted in a steeper LF. 
The detection of image artifacts, such as rings around star residuals, was sensitive to the PSF detection setting as well. 
We finally chose a  PSF-FWHM of 8 pixel equivalent to 7.8 pc (twice the average PSF, the minimum acceptable size for a resolved \hh region) and a high S/N of 10 to completely remove the  false  detections of stars residuals  that were not picked up by either the continuum subtraction or the masking code.  More importantly, this setting prevented  H{\small II}phot from breaking larger regions with some flux fluctuations into false smaller regions. 

It is worth noting that defining the borders of \hh regions in complexes is not a well-defined problem. There is no consensus on whether
 \hh complexes should be considered as giant \hh regions or be broken into smaller regions and if so into what scales.  An automated system such as H{\small II}phot still depends on the aperture definitions but reduces the manual selection bias. In our runs we chose parameters which improved the reliability of the final catalog, at the cost of
some incompleteness at the faint end. The number of missed  regions above the detection limit changes by $10-20\%$ in different fields by selecting S/N=10 instead of S/N=5. The missed regions are responsible for less than $5\%$ of the total detected flux, but considering a higher S/N also affects the region edge  definition and causes differences in measured fluxes up to $20\%$. However, most of the the neglected flux is from DIG and   not to be considered as part of the total H$\alpha$ flux from \hh regions. 

\begin{figure}
\center
\plottwo{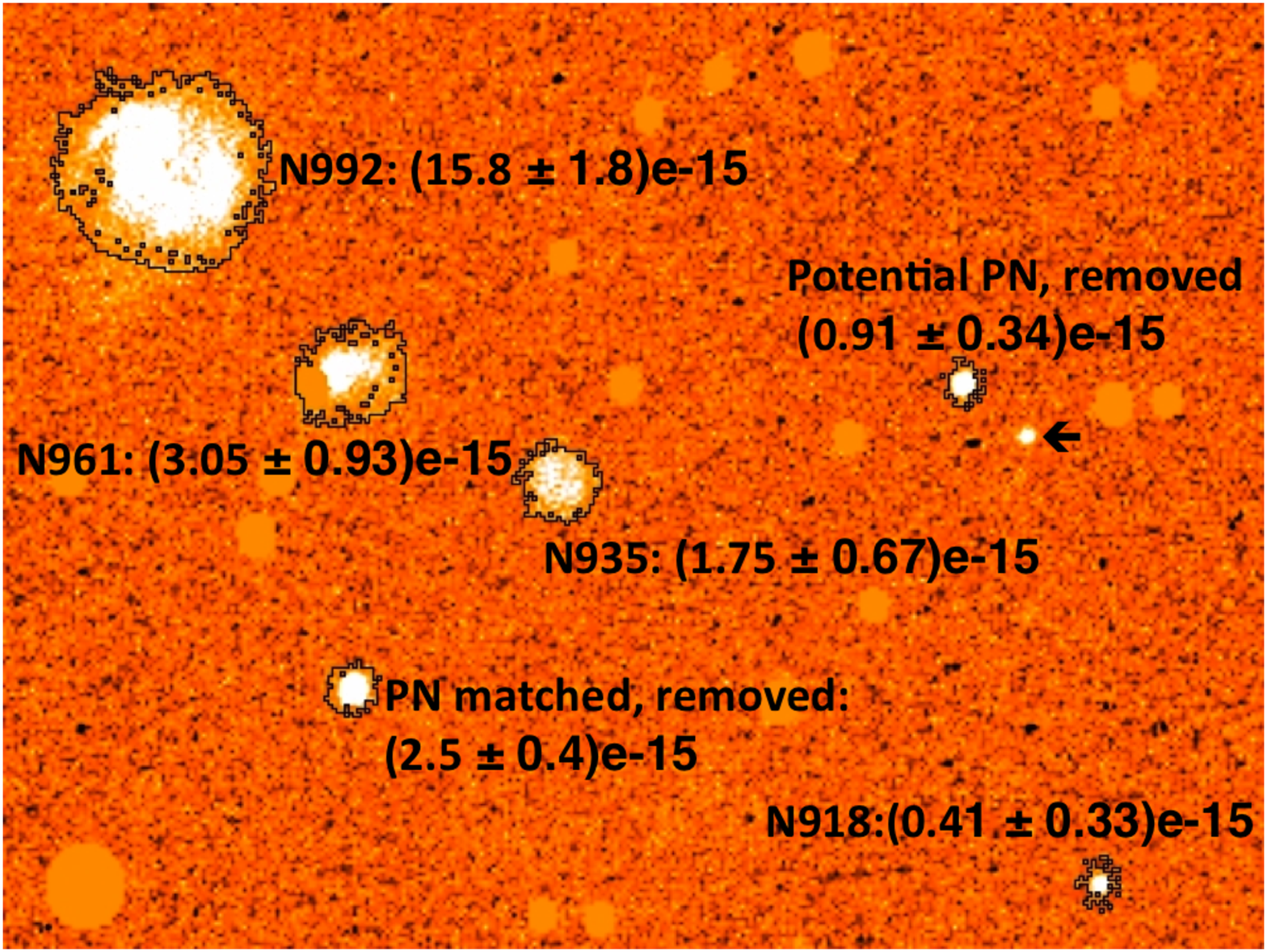}{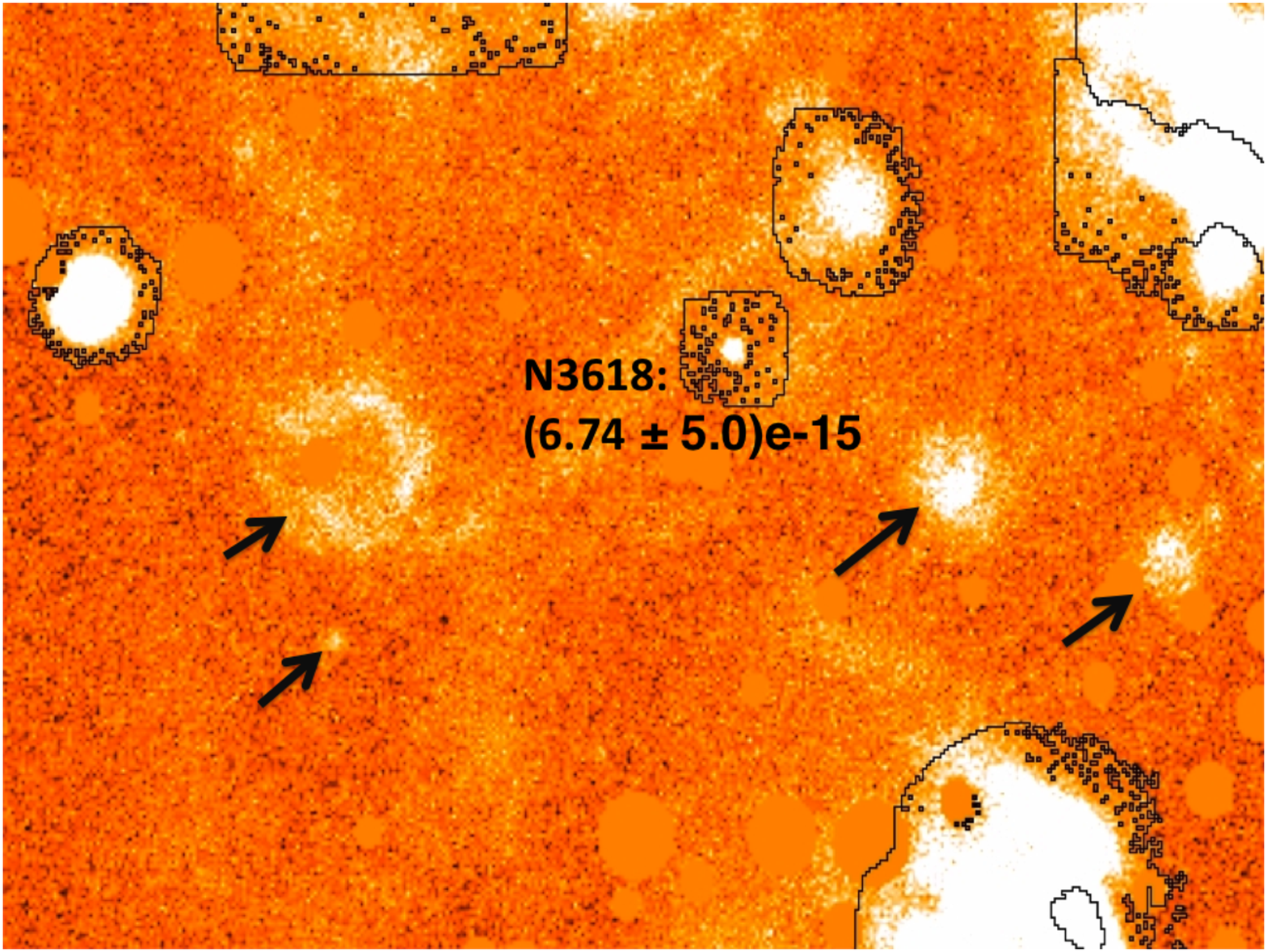}	
\caption{Example of \hh region detection. The solid circles are removed stars. All the regions have an initial S/N detection of 10 or larger, however the final flux uncertainty is relatively large due to the background emission.  The numbers following the source catalog number on each image show the flux and its uncertainty. Two of the detected regions in the left panel were matched with PNe lists and removed. The right panel shows region N3618 which lies within a filament. The source is a compact bright object but due to the DIG background emission, it has a large uncertainty in flux determination. There are four other regions marked by arrows which did not pass the high S/N=10 detection limit. The emission of these regions has been included as part of the DIG in computing the total H$\alpha$ emission of the galaxy.  \label{examples}}
\end{figure}

Background fluctuations and DIG around complexes and in filamentary structures increase the uncertainty in flux measurements.  
We reiterate that there is a  difference between  detection S/N and the flux uncertainty measurement. To calculate the noise we adopt the entire field, rather than the empty regions, as background; this results in a conservative estimate of the noise level. Any source with a signal 10 times larger than this field noise is considered a detection. After growing regions, H{\small II}phot considers the local background in measuring the flux and its uncertainty. As a result the flux uncertainty for regions with high-intensity DIG, large background fluctuations, or regions adjacent to the galaxy center might be noticeably large.   A  large flux uncertainty does not mean a poor detection.  As shown in Figure \ref{examples} from Field 7, all the detected sources are real.   The source number N1017 in the left panel is the one in Table \ref{tb1} with the largest flux uncertainty.  Another source, N1020, also appears to be real but is not counted as a detection. The other source in the bottom left corner has been matched with a planetary nebula and has been removed from the final catalog. In the final catalog we have removed regions with flux uncertainties larger than 80\%. In total only 6\% of the remaining regions have a flux uncertainty larger than 50\% and only 13\% have flux uncertainties larger than 30\%. Almost 75\% of the regions have flux uncertainties smaller than 20\%. 
The right panel of Figure \ref{examples} shows another example of a high flux uncertainty region in Field 4, near the galaxy centre. It lies over an extensive DIG filament, which is why H{\small II}phot was not able to determine an accurate flux. The object is  a real compact source. There are several other extended regions or DIG features in the field marked by arrows, which have not been detected as HII regions.

\section{Final catalog characteristics} 
\label{catalog}

\begin{figure}
  \centering
  \includegraphics[width = 14cm]{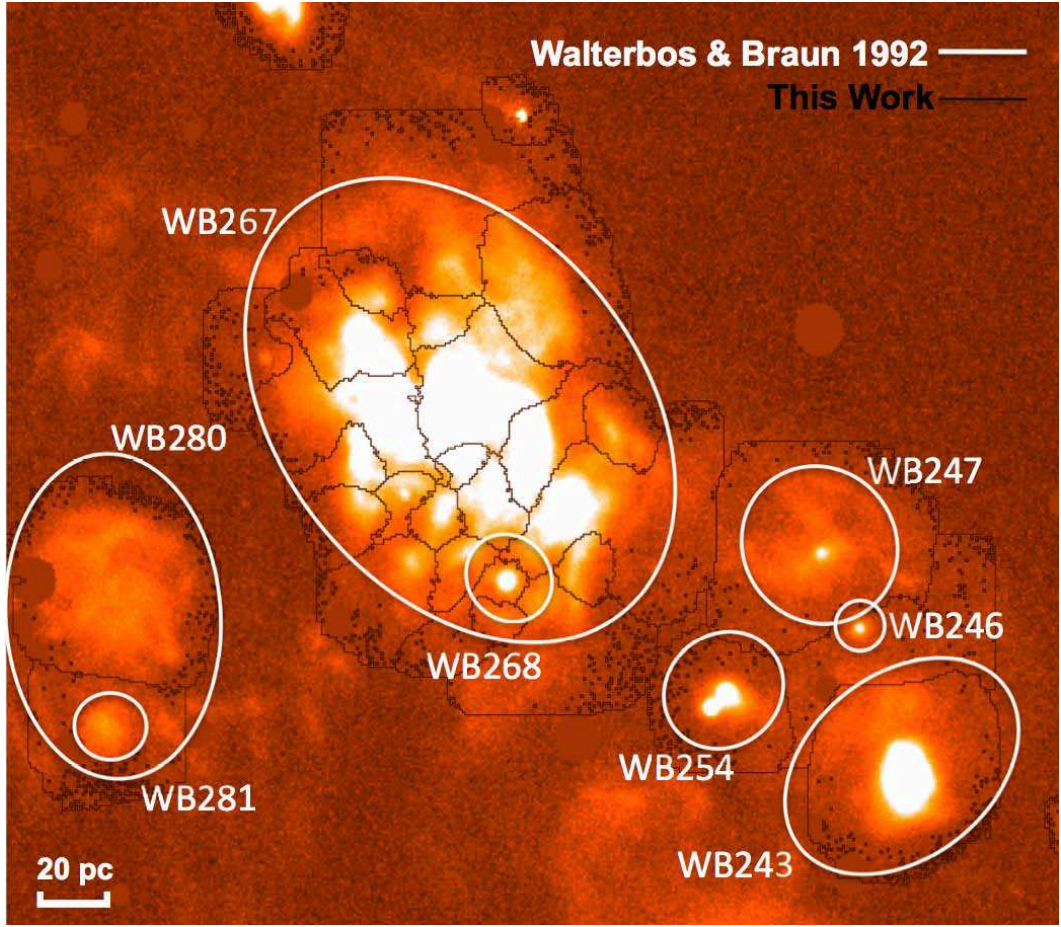}
\caption{A sample comparison between our catalog and WB92. Black lines show the borders determined by H{\small II}phot and white ellipses show the same regions in WB92 (a color version of this figure is available in the online edition).}\label{regions2}
\end{figure}

Our new catalog contains about 5 times as many regions as the catalogs of \citet{pellet78} and WB92.  However, those catalogs contain many regions whose diameters are much larger than the largest ($\sim$190 pc diameter) region from our catalog.  The obvious reason is that our increased  spatial resolution, and use of the HIIphot automated detection method, results in many of their largest regions being resolved as complexes of several smaller regions.  Figure \ref{regions2} shows a sample sub-map, comparing  our  H{\small II}phot detected regions with those of WB92. Black borders are those determined by H{\small II}phot  while WB92 regions are presented with white ellipses. The region WB267 has been divided into 22 distinct regions in our catalog. H{\small II}phot  also avoids  overlap, such as for the regions WB267/WB268 or WB280/WB281, which gives better flux estimates. 

We set a high detection limit of 10$\sigma$ in our survey, yet we can detect H$\alpha$ emission as faint as  $\sim  10^{-16}$ erg cm$^{-2}$  s$^{-1}$ or a luminosity of $10^{34}$~erg~s$^{-1}$ which is  a factor of 5 fainter than  WB92.  With that luminosity detection limit we can easily pick out the B0 stars but cannot go far beyond to B1 stars. 

\hh  regions are not the only H$\alpha$-emitting sources which are detected in our survey. Planetary nebulae (PNe) also emit in H$\alpha$ and may make up a fraction of the detected objects \citep{meysson93}.  \citet{Ciardullo} suggested that there is a natural limit for the [OIII] luminosity of the  brightest PNe.  WB92 used that limit to find the maximum H$\alpha$ luminosity of PNe in M31. Corrected to the new distance for M31, any object which has an H$\alpha$ luminosity smaller than $\sim 5 \times 10^{35}$~erg~s$^{-1}$ could be a PN. To remove the possible PNe detected as \hh regions, we  matched our catalog with  a list of 723 PNe in M31 which covers both disk and bulge of the galaxy \citep{Halliday06}.  We found 374 matches, but 18 of them were brighter than the maximum limit and were not considered.  However the detected flux from these regions is partially contributed by the PNe. The remaining unmatched PNe are either out of our field, too faint to be detected, or removed with star residuals. The matched  and removed PNe were responsible for 1\%  of the  total measured H$\alpha$ emission.

\citet{merrett06}  presented a catalog of 3300 emission-line objects found by the Planetary Nebula Spectrograph in a survey of the Andromeda galaxy. After removing extended objects which are probably \hh regions or background galaxies, 2049 of these objects  covering a large area beyond the galaxy's disk were found to be likely PNe. We matched our PNe-removed catalog with  this list and found 407 common objects which are mostly compact sources. We did not include these objects in our final catalog and analysis but have listed them in Table~\ref{removed}.

The final catalog  contains 3961  \hh regions. It contains about 5 times more regions than WB92 which covered only the northeast half of the galaxy. The luminosity  limit is also five times fainter with L$_{H\alpha}= 10^{34}$~erg~s$^{-1}$.  The average flux  uncertainty is $\sim 5\%$ but it increases in crowded regions with substantial background and especially for faint regions within an extensive background DIG. We removed all the regions with flux uncertainties larger than 80\% from the final catalog.   75\% of the remaining regions have final flux uncertainty less than 20\%  and they are responsible for about 97\% of the total detected flux. 

For each region, the reported position is the coordinates of the flux peak, which is not necessarily at the geometric center of the object especially for  irregular morphologies. The typical peak-to-geometric-center separations range from a few up to 10 arcsec.
A sample of the final results is presented in Table \ref{tb1}. 
The full data  set is  available in the electronic version of the paper and contains the position, dimensions (FWHM major and minor axis in arcsec and full diameter in parsec), position angle,  H$\alpha$ flux, extinction, and extinction-corrected luminosity  of the regions. 

\subsection{Comparison with previous catalogs}
\begin{figure}
\center
\plottwo{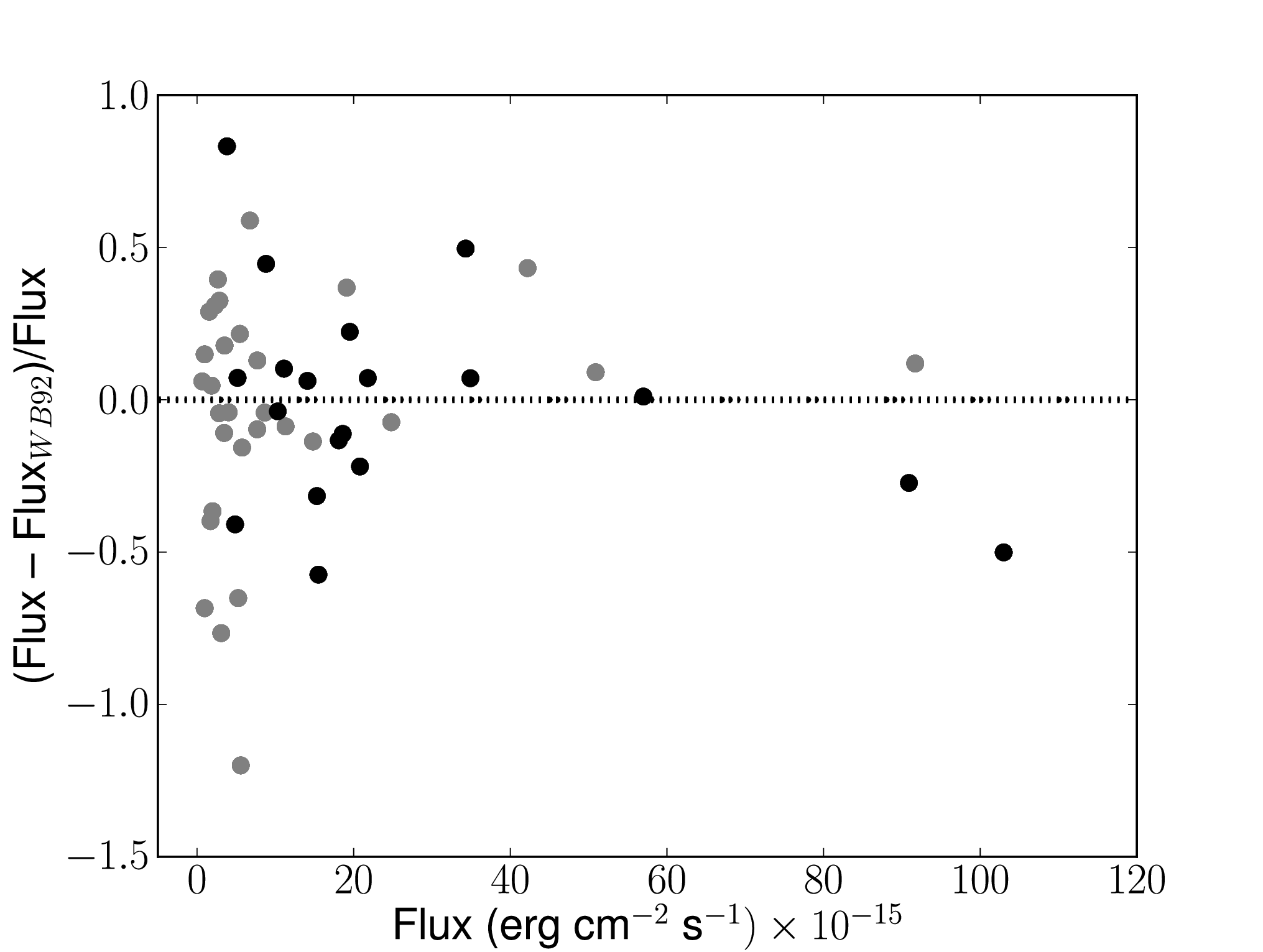}{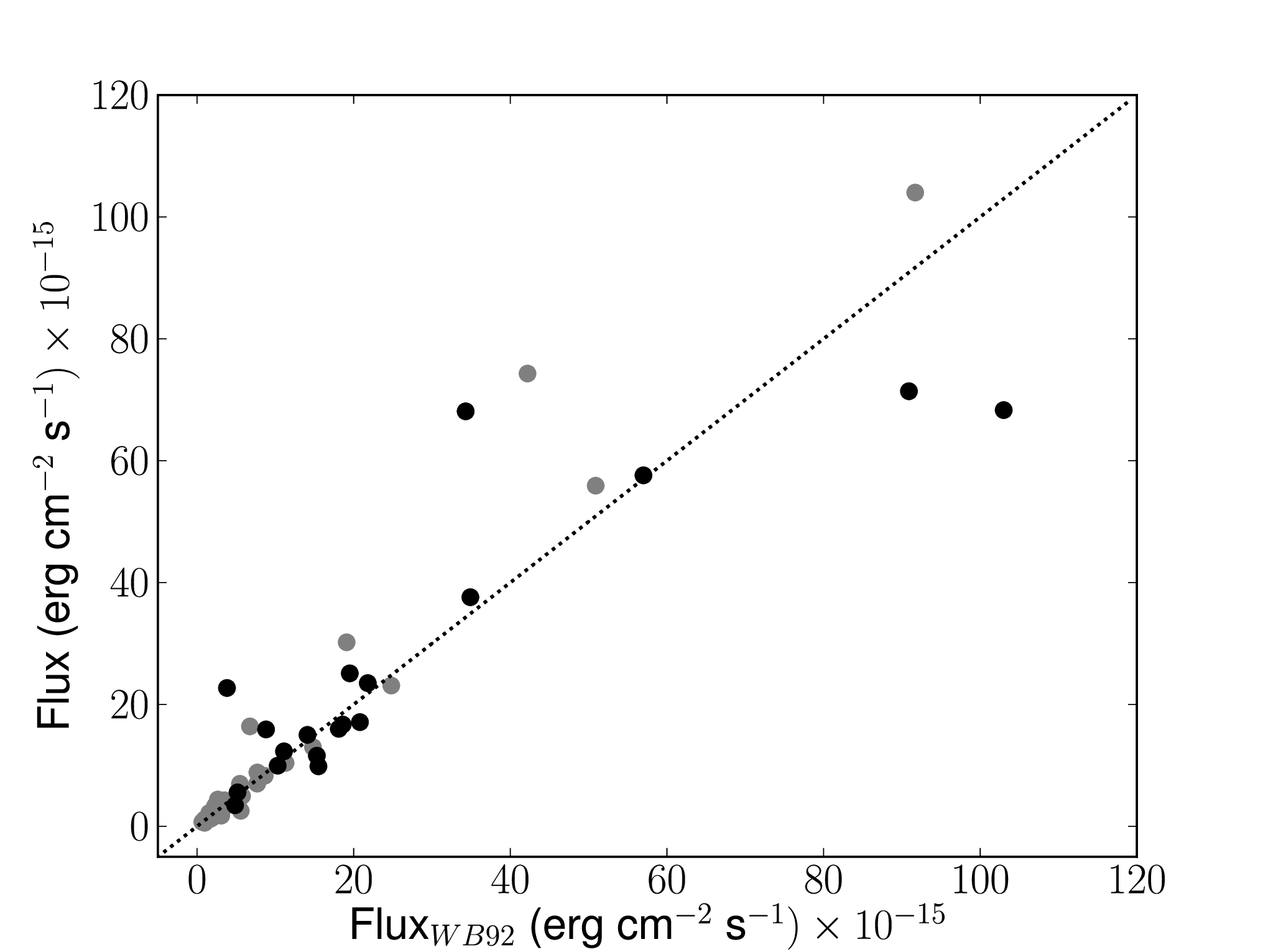}
\caption{Comparison between flux measurements in \citet{walterbos92} and this work. The grey show all the matched regions with a separation $\leq 0.5''$ and black dots  present only larger ($\geq$ 12 pixels, 11.75 pc) well-resolved regions. \label{wbsvsus}}

\end{figure}

We compared our H$\alpha$ fluxes with matched regions from WB92 to check the accuracy of our measurements. To find the best match we considered that many bright complexes detected as one region in WB92 were resolved into smaller fainter regions in our study. Therefore such detections, especially in crowded regions which made the comparison of total fluxes complicated, were excluded. 
There were also plenty of  individual faint regions matched in both catalogs but they were under or close to the  resolution limit with larger uncertainties in flux determination and were excluded as well. Finally we found 49 regions well matched in coordinate space, not contaminated by neighbours due to poor resolution, and luminous enough to be well above the detection limit. Figure \ref{wbsvsus} shows our measured fluxes versus WB92. The grey dots show all the matched regions with a separation $\leq 0.5$\arcsec\ and black dots  present only extended ($\geq$ 12 pixels, 11.75 pc) well resolved regions. There is more scatter for larger  fluxes but an average ratio of  $1.00\pm0.12$ is very satisfying. Most of the scatter for larger fluxes results from the difference in edge finding and background correction especially in crowded regions in the spiral arms.

The fact that  M31  does not  have very luminous \hh regions such as 30 Doradus \citep[$\log(L_{H\alpha})=40.17$;][]{kennicutt84} in the LMC and NGC604 
\citep[$\log(L_{H\alpha})= 39.49$;][]{relano09} in M33 was addressed before (KEH89, WB92). These two regions are both excited by young massive star clusters   containing hundreds of exciting stars and are the largest \hh regions in the Local Group. The brightest detected \hh region in our study has a luminosity of  $\log({\rm L}_{H\alpha})=37.8$ before extinction correction  which is smaller than the cut offs detected by KEH89 ($\log({\rm L}_{H\alpha})=38.6$) and WB92  ($\log({\rm L}_{H\alpha})=38.2$). Their catalogs have problems resolving faint regions especially within crowded spiral arms and at the borders of the brighter \hh regions. Our higher detection sensitivity  has  resolved such regions into smaller ones or individual \hh regions (Figure \ref{regions2}). We discuss the resolution effect in more detail in section \ref{sizefunc}.

\section{Analysis and Discussion} 
\label{discussion}

\subsection{\hh Region Luminosity Function in M31}
 \label{LF}

Studying the \hh region LF is very important for examining the distribution of massive stars and therefore the star formation  in galaxies. The stellar  IMF, which is one of the fundamental  properties of the star formation process, is  believed to be universal as first derived by \citet{salpeter55}, but new studies suggest a dependence on environmental conditions  \citep[e.g.][]{krumholz08, meurer09}. 
Massive stars are the only components that can be well-observed  in other galaxies to examine the universality of the IMF, although only at the
high-mass end \citep[e.g.][and references therein]{calzetti2010}. 
\hh region LFs  in nearby galaxies have been extensively studied  as a characteristic parameter  of galaxies  \citep[e.g.][]{kennicutt88,banfi93,rozas96,knapen98,feinstein97,oey98,thilk2000,thilk02,oey03,gutier11}. \citet{kennicutt88} and KEH89  investigated the \hh region populations in different types of galaxies and found a large variation along the Hubble sequence. In general, the total number of \hh regions increases significantly from Sb galaxies  to  later Hubble types. The early type galaxies also have a steeper LF and a significant drop in number of very luminous regions with ${\rm L}_{H\alpha} > 10^{39}$ erg s$^{-1}$. \citet{gonzalez97} studied a sample of spiral galaxies (S0 to Sbc) with active nuclei but they did not find any dependence of the \hh LF with  Hubble type of the galaxy.

The  observed H$\alpha$ emission of \hh regions is partially absorbed by dust in the galaxy's interstellar medium. As a result, the measured flux and luminosity are underestimated, especially toward dense  spiral arms. 
We used the dust optical depth at H$\alpha$ wavelength map made by \citet{Taba2010} based on the {\em Spitzer}  MIPS data \citep{gordon2006} to correct the luminosities. More than 80\% of the detected regions were covered in the extinction map. For the remaining 20\% we used the uncorrected luminosity but most of these regions are located within the outer disk and are less likely to be  affected by dust absorption (Figure \ref{inarms}).  
The pixel size of the extinction map is 60--70 times  larger than that of our images, therefore we matched the position of the peak at each region with the corresponding pixel in the extinction map  and used that number  to correct the total luminosity of each region as:
  \begin{equation}
 2.5  \log(\frac{L_{corr}}{L_{obs}})= A_{H\alpha}
 \end{equation}
where we have used $A_{H\alpha} = 1.086 \tau_{H\alpha}$ \citep{caplandeh86}.

\begin{figure}
\center
\includegraphics[width = 12cm]{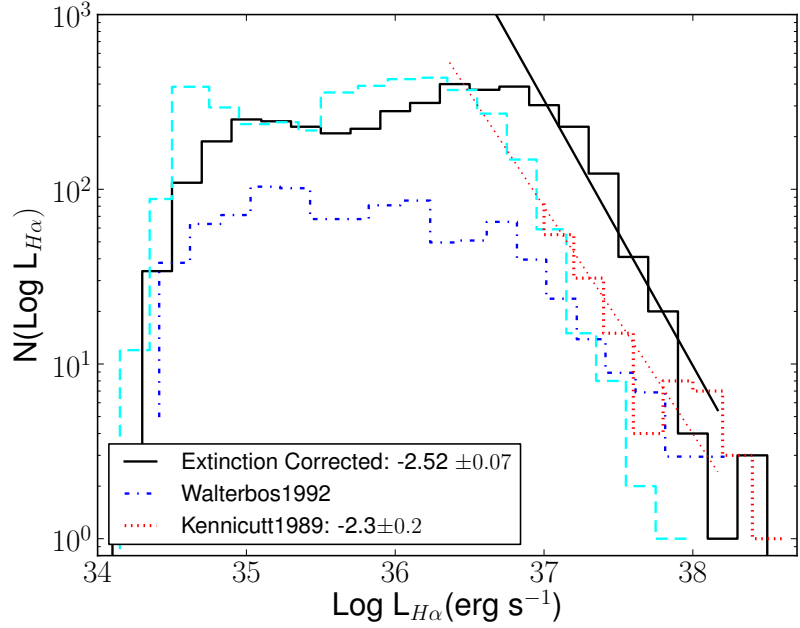}
\caption{ M31 \hh region luminosity function, after removing the planetary nebulae. The dashed cyan and  solid black histograms show the LF before and after correction for extinction. 
The solid black line shows  the least squares fit with slope of $\alpha = -2.52\pm 0.07$ using an average of  four different 0.2~dex bins within  $\log({\rm L}_{H\alpha}) = 36.70-37.80$ each shifted 0.05~dex. For comparison the KEH89 and the WB92 results are plotted in red dotted  and blue dash-dotted lines respectively.  The dotted red straight line shows the KEH89 power law with $\alpha = -2.3\pm0.2$ [note that the fitted slope for $\log N (\log {\rm L}_{H\alpha})$ versus  $\log ({\rm L}_{H\alpha})$ is $1-\alpha$]. Two distinct  peaks  at luminosities corresponding to B0 and O7--O9 stars are noticeable.  \label{lumfunc}}
\end{figure}

\begin{figure}
\hspace{-1cm}
\center
\includegraphics[width=14cm]{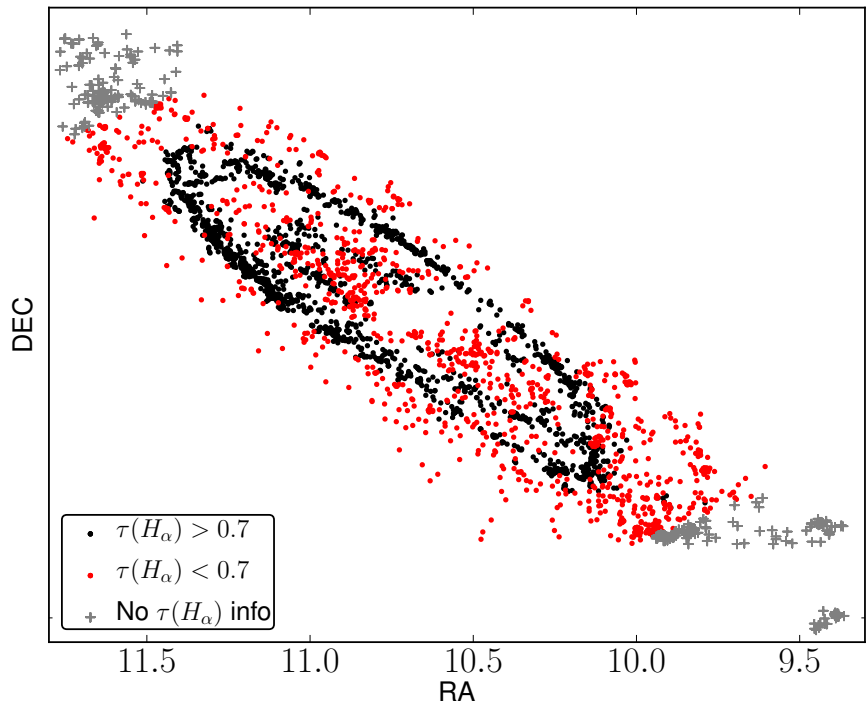}
\caption{
In-arm and inter-arm regions selected based on the spiral arm structure in dust.  Black and red dots present regions with $\tau$(H$\alpha$)$> 0.7$ and  $\tau$(H$\alpha$)$< 0.7$ respectively. Grey crosses are regions for which we do not have the extinction information, but based on their location at the edges of the disk, we have considered them as inter-arm regions. 
\label{inarms}}
\end{figure}

The primary LF and the final extinction-corrected LF are plotted in Figure \ref{lumfunc}.  We have plotted the LFs derived by KEH89 and WB92 for comparison. 
The LF is flat in the middle with sharp declines at both bright and faint ends, and is  qualitatively similar to those
derived in previous works.
Turnover points are defined as where the bin counts begin to steadily decrease. KEH89 found that for most galaxies in their sample, the turnover point occurred between $36<\log(L_{H\alpha})<37$. However  their  data (Figure \ref{lumfunc}) were not deep and resolved enough to find a turnover point for M31. The LF of WB92 shows  a flattening between $\log(L_{H\alpha})$=35.2 and 36.8  and a power-law index of $-1.95$ for  
$36.8< \log(L_{H\alpha})<38.2$. WB92 also suggest a flat theoretical extrapolation for  \hh LF by KEH98 for $\log(L_{H\alpha}) < 37$.

A  turnover point in the \hh LF  is not a characteristic of every galaxy, especially for irregulars  \citep{hodge90, hodgelee90}. 
\citet{feinstein97} predicted that a flat LF or turnover point could be modelled by considering a constant star formation rate, but a complicated star formation history, such as a burst in the past, would form more faint objects than a constant SFR and make the \hh LF complicated. 
\citet{youngb99} found that their sample of 29 normal Im galaxies could be divided into two categories: ``Turnover" and ``No Turnover". They argued that the turnover point was not a completeness problem at the low end of the LF, as suggested in some previous works.  Galaxies with ``Turnover" LFs were found to have a larger number of \hh regions, larger luminosity  cut-offs, steeper power-law fits and higher SFR per unit area.  

Figure~\ref{lumfunc} clearly shows that the \hh region LF in M31 has a faint-end turnover.
We are resolving most of the very  luminous regions ($\log(L_{H\alpha}) > $37.5),  ionized by multiple  stars, into smaller regions and even individual stars.  Therefore we adopted a smaller turnover point of $\log({\rm L}_{H\alpha})=36.7$,  and the same bin size of 0.2 dex, to be able to compare the results with previous works and to achieve the best fit.
The LF was fit using a standard  power law:
\begin{equation} \label{LF-eqn}
N(L) dL= AL^{\alpha} dL .
\end{equation}
We used an average of four  0.2 dex bins  with shifts of  0.05 dex  in the range   $\log({\rm L}_{H\alpha})=36.7-37.8$ to get better statistics and the best correlation coefficient ($r^{2}=0.96$ and $\chi^{2}=2.5$). 

A power-law index  of $\alpha = - 2.52\pm 0.07$ was obtained  using least squares fitting and is plotted as the solid black line in  Figure \ref{lumfunc}. This slope is steeper than but consistent with  the $\alpha = - 2.3\pm0.2$ reported by KEH89 and plotted as the red dotted line in Figure \ref{lumfunc}. We are resolving most of the luminous regions in the bright end of the KEH89 LF (which refers to multiple ionizing stars), into smaller regions and even individual stars.  Therefore a steeper LF, especially at the bright end,  is expected. The derived fit, however, is highly dependent on the range of luminosities used. 

A double power-law for $38<\log({\rm L}_{H\alpha})<39$, known as a type~II LF, has been reported for individual galaxies in many \hh LF studies 
\citep[e.g.][]{kennicutt89, rand92,rozas96,rozas99,thilk2000,gutier11}. \citet{bradley06} used an ${H\alpha}$ imaging survey of 53 nearby  galaxies \citep{knapen04} to make  a composite  LF of 17797 \hh regions. They found that the LF is steeper for larger luminosities and breaks at a luminosity of 
$\log({\rm L}_{H\alpha})=38.6\pm0.1$ with a sharp fall for  $\log({\rm L}_{H\alpha})>40$. \citet{pleuss2000} discussed the resolution effect on determining the \hh LF and argued that the type II LF might happen because of the overlapping and blending of smaller \hh regions leading to higher measured luminosities. 
The modelling  of \hh LFs by  \citet{beckman2000} showed that in case of clustering or overlapping, the LF slope at higher luminosities should decrease,  not increase. Instead they suggested  that the ``glitch'' might be caused by the transition  of \hh regions from ionization bounding at low luminosities (at a critical mass where the \hh region only ionizes its cloud)  to density bounding (in which the larger flux of Lyman continuum photons ionizes the diffuse gas and even the intergalactic medium).  

The M31 LF upper cutoff is below the type~II LF break ($\log({\rm L}_{H\alpha})=38.6$),  which would, in the interpretation of \citet{beckman2000}, imply that all M31 \hh regions are ionization bound. However, a large fraction of the total  measured ${H\alpha}$ luminosity in M31 is emitted by DIG. \citet{giammanco04} showed that optically thick clumpy models for \hh regions allow a significant fraction of the ionizing photons emitted by the exciting  stars to escape from their \hh regions. Our observations (e.g. Figure \ref{regions2}) confirm the highly clumped structure of the ${H\alpha}$ emission. One of the main reasons that we adopted a larger minimum acceptable size than the image resolution to identify the \hh regions was to avoid false detections due to such fluctuations (e.g., see Figure \ref{examples}).

A large fraction of the detected regions are smaller than the $\approx 10^{37}$~erg~s$^{-1}$ transition point between \hh regions ionized by a single star and those ionized by associations and clusters. Therefore a large fraction of detected regions in our study are comparable to typical Galactic \hh regions ionized by single OB stars. It is important to note that most of the H$\alpha$ emission studies in nearby galaxies only probe the  massive-star-formation-containing complexes and clusters. 
\hh regions fainter than $\sim 10^{37}$~erg~s$^{-1}$ have not been well studied in external galaxies. They were  assumed to generate only a small fraction (5--10\%) of the total 
H$\alpha$ emission of the galaxies (e.g. KEH89, WB92). M31 is the only exception: KEH89 and WB92 estimated that faint regions contributed
at least  30\% of the total H$\alpha$  luminosity. In our survey  we have well resolved the major fraction of individual OB stars in associations and field regions below this limit. In total  they are responsible for more than  40\% of the detected source emission in M31. This result is in contrast with the \hh region population and emission in irregular galaxies. \citet{youngb99} found that for most of their Im galaxies, 80\% of the H II region luminosity is emitted from  complexes
with typical luminosities of $10^{37}-10^{38}$ erg~s$^{-1}$,  comparable to 10 Orion nebulae.

Besides the number of faint \hh regions in M31, the other notable feature of Figure \ref{lumfunc}
is that the LF appears to be double-peaked. Our data are complete to L$_{H\alpha}$=10$^{34}$ erg~s$^{-1}$, 
therefore a second peak at $10^{35}$ erg~s$^{-1}$ suggests a second population of stars. The true location of the second peak may be affected by incompleteness.  
The first peak at $10^{35}$ erg~s$^{-1}$ and the second peak at $8\times10^{36}$ erg~s$^{-1}$ correspond to 6.6$\times 10^{46}$ and $5.2\times10^{48}$ Lyman-$\alpha$ photons~s$^{-1}$ which indicates the emission sources are B0--B1 and O7--O8 stars respectively \citep{spitzer78}.  
The peak at B star luminosities suggest an aged population of massive stars which has lost most of its O stars. The second peak at O stars' luminosities shows the current massive star formation in M31.  
It also matches the transition luminosity between \hh regions created by individual stars and complexes.
The O stars mostly lie in the spiral arms while the B stars fill the inter-arm regions as well. We discuss the in-arm/inter-arm distribution of the two populations in more detail in section \ref{arms};  
however the fact that a fraction of these faint regions are compact and are distributed all over the galaxy disk means that they might be unidentified PNe and not really B stars. 

According to \citet{feinstein97}, a past starburst might be the cause of the first peak at  L$_{H\alpha}$=10$^{35}$ erg~s$^{-1}$. Comparing the lifetimes of O 
and B stars ($\sim 15\times 10^{6}$~Myr for an O8 star, $\sim 20\times 10^{6}$~Myr for  a B0.5 star) suggests that  M31 experienced a star burst between 15 and 20 million years ago. The UV study of the history of star formation in M31 by \citet{kang09}  also confirms a recent peak in star formation between 10--100 million years ago.
The timescale derived here for a recent starburst in M31  is an order of magnitude smaller than  the 210 Myr dynamical time derived by \citet{gordon2006} for   the  head-on collision between M31 and M32 \citep{block2006} that created the 10 kpc star forming ring in M31. \citet{mcconnachie09}  detected stars  that were  remnants of dwarf galaxies destroyed by the tidal field of M31 and suggested  that 75\% of M31's dwarf satellites are not yet known.
We suggest that the double-peaked LF might also be the result of a more recent collision of M31 with a dwarf satellite galaxy.

\subsection{Total {H$\alpha$} luminosity and star formation rate}

M31's \hh regions are not very luminous compared to giant complexes in other Local Group galaxies. 
We calculated the total H$\alpha$ luminosity from M31 \hh regions  to be $1.77\times10^{40}$ erg s$^{-1}$, after  extinction correction  by an average factor of 2.72.  This luminosity is comparable with the 30 Doradus complex in the LMC with L$_{H\alpha}=1.5 \times 10^{40}$~erg s$^{-1}$ \citep{kennicutt84}.
It does not include the emission from DIG and  PNe. To estimate the contribution to the total galaxy emission from DIG and undetected regions,  we measured the total integrated luminosity from each field and subtracted the sum of the \hh regions' luminosities. 
On average we determined a $\sim 65\%$ contribution from DIG, PNe and undetected sources. Removed PNe comprise less than 2\% of the total luminosity which is even less than the flux measurement uncertainty.

\citet{walterbos94} also reported a large DIG contribution (40\%) for M31. Spectrometry studies \citep{greenaw97,galarza99}  confirmed that the DIG in the disk of M31 is photoionized by a dilute radiation field and there is a smooth transition for line ratios  between \hh regions and DIG in M31.  \citet{thilk02} mentioned  a leak of ionizing photons from \hh regions to the DIG which causes a larger contribution of L$_{H\alpha}$ by DIG. 
\citet{beckman2000} modelled high luminosity \hh regions in disk galaxies and showed that regions with L$_{H\alpha}>10^{38.6}$ might be density bounded. Such luminous  regions emit a tremendous amount of  Lyman continuum photons, sufficient not only   to totally  ionize the surrounding cloud but also to  ``leak'' into the the intergalactic medium. All the \hh regions in M31 are smaller than density-bound limit, yet a large fraction of the total H$\alpha$ emission is contributed by DIG in the disk of the galaxy itself. 
In another scenario, \citet{giammanco04} suggest that the clumpy structure of \hh regions may  help  a large amount of  ionizing photons to leak into the surrounding clouds beyond the boundaries  of \hh regions.

The  masked and partially saturated $10'\times10'$ bright bulge region  also contains a maximum of 10\% of the total H$\alpha$ luminosity. 
To estimate this contribution we measured the ratio of the  total counts in the $10'\times10'$  central region to the total counts in the F5-H$\alpha$ field which contained  the bulge region. To estimate the total bulge contribution we assumed that every pixel in the $\sim 2.5\arcmin \times1.5\arcmin $ masked bulge in the original image had the limiting saturation value. 
Then the total luminosity of this field was compared to the total luminosity.
Considering that upper limit we calculate a maximum H$\alpha$ total luminosity of 5.6$\times10^{40}$ erg s$^{-1}$ for M31. The major uncertainty is caused by the dust attenuation  estimation \citep[$\sim10\%$, cf.][]{Taba2010} and the DIG estimation (10-15\% due to object detection uncertainty in defining DIG). Adding the uncertainty in distance and flux estimation, we have a total uncertainty of $15-20\%$ in the total H$\alpha$ luminosity.  Our value is consistent with  L$_{H\alpha}=4.1 \times$10$^{40}$~erg~s$^{-1}$ [$4.2 \times$10$^{40}$ corrected for distance and using dust attenuation of 2.72 instead of 3.4) determined by \citet{walterbos94}].

\citet{devereux94} reported a luminosity of  (2.8$\pm$0.88)$\times$10$^{40}$ erg s$^{-1}$ [$(6.2\pm 1.9) \times$10$^{40}$ erg s$^{-1}$, corrected for distance, our own extinction factor and 35\% NII contribution]. 
Their estimation also suffered from background continuum subtraction which led to a 50\% flux uncertainty in some filamentary interior regions.
Determining the total luminosity is very uncertain due to different assumptions about how to correct the observed flux.  For example \citet{barmby06} and \citet{Taba2010} used the raw total H$\alpha$ luminosity measured by \citet{devereux94} and  derived two different values of  9.98$\times10^{40}$  and 4.75$\times10^{40}$ erg~s$^{-1}$ for corrected L$_{H\alpha}$, considering different values for extinction and the [NII] contribution. 

We can compare the total H$\alpha$ luminosity of M31 with other estimates of its overall star formation rate. M31 has been described as a quiescent galaxy with low star formation \citep{kennicutt88}. 
Using the IRAC 8~$\mu$m non-stellar luminosity density,  \citet{barmby06} determined a SFR of $0.4$~M$_\odot$ yr$^{-1}$. We converted the H$\alpha$ luminosity to SFR using the relation of SFR [M$_\odot$~yr$^{-1}]=7.9\times10^{-42}$~L$_{H\alpha}$~[erg s$^{-1}$] \citep{kennicutt98} and derived a value of
$0.44$~M$_\odot$ yr$^{-1}$, close to that of  \citet{barmby06}. 
\citet{kennicutt09} suggested that a combination of IR and H$\alpha$ gives a better dust-corrected estimation for SFR. We followed their calibration as:
\begin{equation}
{\rm SFR(M)}_{\odot} {\rm yr^{-1}} = 7.9\times10^{-42}[{\rm L}(H\alpha)_{\rm obs} + 0.020{\rm L}(24)]({\rm erg s}^{-1})
\end{equation}
where ${\rm L}(H\alpha)_{\rm obs}$ is the H$\alpha$ luminosity without correction for internal dust attenuation and ${\rm L}(24)$ is the 24~$\mu$m IR luminosity. Adopting  $L(24)=1.09 \times 10^{42}$ erg s$^{-1}$ from MIPS 24~$\mu$m observations
\citep{montalto09}, the above equation gives a present day SFR of 0.34~M$_\odot$~yr$^{-1}$ which is consistent with  the \citet{Taba2010} range of 0.27--0.38 M$_\odot$~yr$^{-1}$.   A comparison of different M31 SFRs from previous studies is summarized in Table \ref{SFR}. They cover a wide range from 0.27 to 1 M$_\odot$~yr$^{-1}$. In a  UV study on star forming regions in M31, \citet{kang09}  argued that M31 has no FUV-detected massive star forming regions younger than 50 Myr. Those authors concluded that  the SFR in M31 is decreasing from a possible peak between $\sim10$ and 100 Myr ago. 

\subsection{Comparison of In-arm and Inter-arm Regions}
\label{arms}

Statistical studies of \hh regions often show differences in properties between sources located in galaxies' spiral arms or in the gaps between the arms 
\citep[e.g.][]{rand92,thilk2000,scoville01}. For example, Figure \ref{regions1}  shows that the bright sources in M31 are generally located only in the arms and noticeably  trace the 10~kpc star forming ring while the fainter sources are scattered between the arms as well. \citet{banfi93} reported a morphology dependence of the \hh LF and also different \hh LFs for in-arm and inter-arm regions in a sample of Virgo cluster galaxies.  \citet{rand92} and \citet{thilk2000} also found different LF slopes for in-arm and inter-arm regions but the difference was not significant. Different LFs for in-arm/inter-arm regions  have not been observed in many other spiral galaxies  \citep[e.g.][]{knapen98,rozas96,gutier11}. The presence of larger \hh regions within spiral arms might be only a statistical effect, not a physical difference between in-arm and inter-arm regions: spiral arms contain more \hh regions and therefore  also more luminous regions \citep{knapen98}. 

\citet{oey98} used Monte Carlo simulations to study the evolution of LFs and suggested that the arm populations represent the currently active star forming regions while the inter-arm regions are aged populations. However, those authors noted that the timescale of passing between spiral density waves at any location within the disk is about 40 Myr which is much larger than the maximum observable 24 Myr lifetime of the \hh regions \citep{oey03}. Therefore a  fraction of new stars must be born in inter-arm regions as well. 

\begin{figure}
\hspace{-1cm}
\center
\includegraphics[width=12cm]{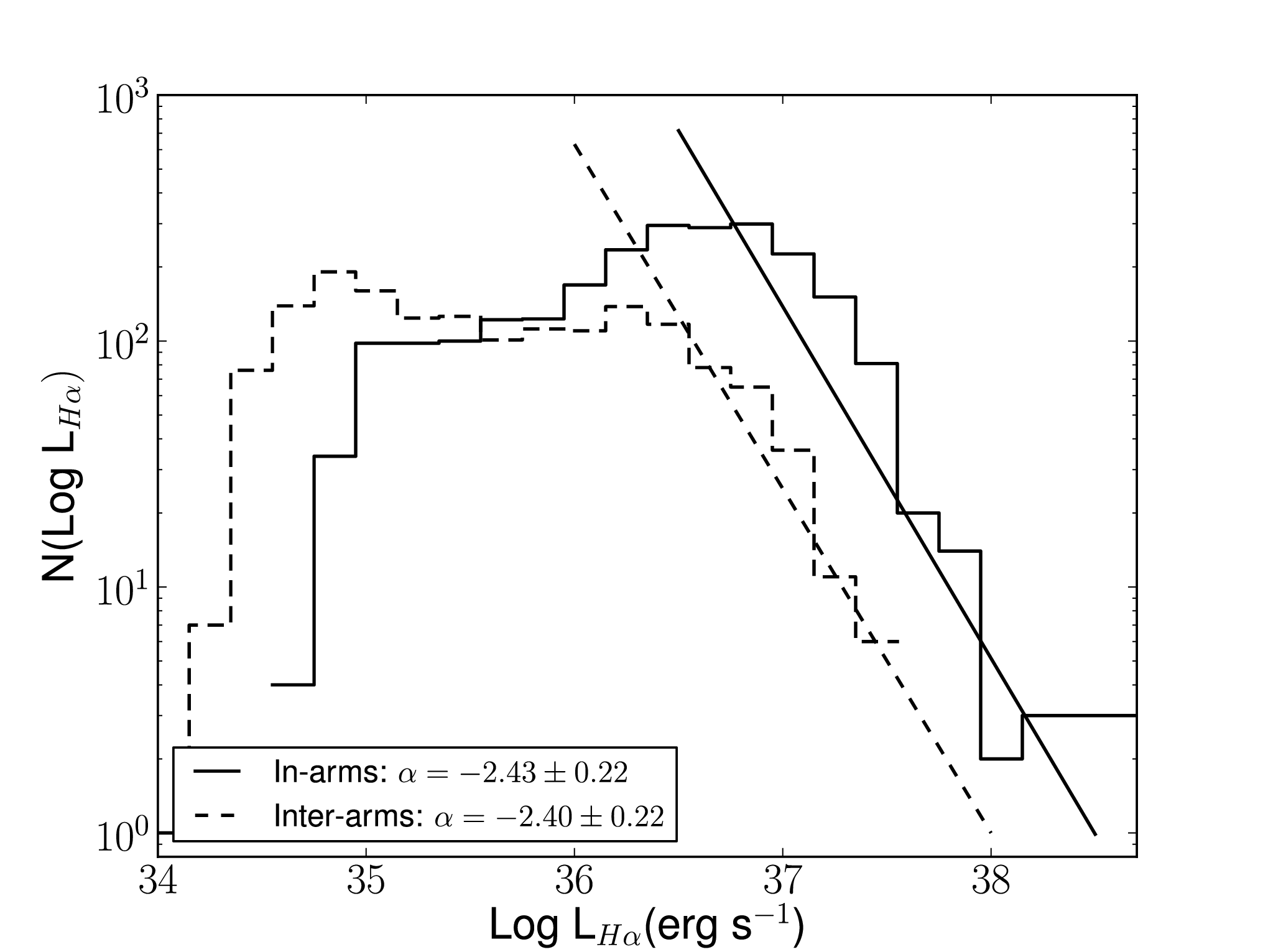}
\caption{ The in-arm/inter-arm luminosity functions selected by H$\alpha$ optical depth.
 It is noticeable that brighter regions lie within the arms while fainter regions cover the inter-arms.  The spiral arms contain most of the young newborn stars and, similar to the total, their LF peaks at O6-O7 stars. As suggested by \citet{oey98} the aged  stars may leave the original cloud in which they form and fill up the gaps between the arms. 
 There is no significant difference in the power law slopes for the two categories. 
 \label{inarmlf}}
\end{figure}

The spiral arms in M31 are identified by highly  concentrated  gas and dust.   \citet{Taba2010} showed a good correlation between the H$\alpha$  and 70 $\mu$m emission; therefore the distribution of the dust optical depth at H$\alpha$ wavelength is an appropriate mask to select in-arm and inter-arm  \hh regions. We used the \citet{Taba2010} $\tau_{H\alpha}$  to designate the arm \hh regions. All the sources with extinction $> 0.7$ were assigned to the arm population. 
Figure \ref{inarmlf} shows the in-arm and inter-arm LFs using the dust distribution masking. The solid line and dashed lines present in-arm and inter-arm regions respectively. The straight lines are the power law fits. 
The in-arm regions are best fitted with  a power-law with $\alpha=-2.43\pm0.22$  and inter-arm regions with $\alpha=-2.40\pm0.22$. 
The inter-arm regions are less luminous and less populated and comprise only 40\% of the total detected regions and 14\% of the total detected emission.  The brightest inter-arm region has a luminosity of ${\rm L}_{H\alpha}=2.80\times10^{37}$~erg~s$^{-1}$, nearly 10 times fainter than the most luminous object with  ${\rm L}_{H\alpha}=2.21\times10^{38}$~erg~s$^{-1}$.

\citet{knapen98} suggested that the lack of bright \hh regions in inter-arm regions is only a statistical effect and that the same LF slope for in-arm and inter-arm regions confirms that both have the same population. However their data were not deep enough to detect individual faint \hh regions created by aged B stars.
Figure \ref{inarmlf} clearly shows two different populations  for in-arm and inter-arm regions in M31. The in-arm LF peaks at brighter regions and presents current star formation. The inter-arm LF peaks at  aged B stars, as suggested by \citet{oey98} and \citet{oey03}. If we accept that inter-arm regions have the same population as in-arm regions but scaled to smaller numbers, then we also expect fewer \hh regions in lower luminosity bins. Surprisingly, the number of \hh regions created by B stars between the arms is noticeably larger than the in-arm regions within the same luminosity bin.  As discussed before the dynamical travel time between spiral density waves ($\sim$ 40 Myr) is larger than the evolution time of \hh regions ($\sim$ 24 Myr).  Therefore to explain the extraordinary large number of faint \hh regions between M31's spiral arms we should either assume that a large number of massive stars are formed in inter-arm regions (which is not probable because there is not enough gas to form too many massive stars) or accept that the SFR has been larger in the past. 
The noticeable fraction of B stars in the inter-arm regions therefore supports the hypothesis of a recent star burst in M31.

\subsection{Size Distribution of \hh Regions in M31} 
\label{sizefunc}

In contrast to the Milky Way, LMC, or many spiral  galaxies such as M51 and M33, M31 does not have very luminous \hh regions. Here we examine whether this is due to higher resolution in M31, resolving larger regions into smaller ones, or an intrinsic lack of giant \hh regions in M31. 

\citet{scoville01} studied the \hh regions in M51 with high angular resolution (0.1\arcsec--0.2\arcsec, 4.6--9.3 pc) {\em HST} imaging. With spatial resolution similar to this work, they detected \hh regions between 10--250 pc with L$_{H\alpha}\approx2\times10^{36}-2\times10^{39}$~erg~s$^{-1}$. 
\citet{gutier11} made a similar {\em HST} survey of M51 and found a size range of 8.3--530 pc. They found two populations which are separated at D$\simeq$120 pc and suggested that size as a transition between \hh regions ionized by a single OB cluster and multiple regions. 

Both  groups defined the effective size based on the area covered by the region rather than the FWHM as used in this work. The FWHM  size is slightly smaller especially in diffuse regions but the size range of M31's \hh regions ($\sim16-190$ pc) is still comparable with M51 and the Galactic \hh regions. Despite the similarity in size, M31 does not have very luminous \hh regions like M51 or the most luminous Galactic   \hh regions such as W49  (L$_{H\alpha}=9.1\times10^{38}$~erg~s$^{-1}$) and W51 \citep[L$_{H\alpha}=3.9\times10^{38}$~erg~s$^{-1}$;][]{schraml69}.  A similar {\em HST} study by \citet{pleuss2000} on M101's \hh regions showed the same range of size (10--220 pc) and luminosity ($\sim10^{36}-2\times10^{39}$~erg~s$^{-1}$). 

Assuming a lifetime of $3\times10^{6}$~yr  for the OB stars to produce most of the ionizing photons, \citet{scoville01} considered an upper limit of 50 pc for the radius of \hh regions created by a single OB star cluster.  They concluded a maximum luminosity of L$_{H\alpha}\sim10^{39}$~erg~s$^{-1}$ for a single cluster and reported any larger or brighter region as a blend of multiple regions. All the regions in our catalog have smaller luminosities  than this upper limit, but there are some regions extended to radii larger than 50 pc. \hh regions grow very fast in their early  phase until they reach the Str\"{o}mgren radius. The ionization front may expand to the diffuse ionized gas after a steady hydrodynamic growth beyond the Str\"{o}mgren sphere. Extended \hh regions in M31 are most probably evolved regions which are not as luminous as when they formed.  They might look extended due to the leakage of the ionizing UV photons as well. The lack of giant \hh regions in  M31 remains as yet unexplained.

The power law LF of \hh regions is well accepted for all different types of galaxies, although the slope apparently depends on galaxy type and may vary for different galaxy components. One then expects to observe a power-law size distribution as well, because the luminosity depends on the volume and therefore the size of the \hh region \citep{bergh, oey03}. 
Assuming $L\propto D^3$ and $dL/dD\propto D^{2}$, the  differential size distribution should be \citep{oey03}:
\begin{equation}
N(D)dD=N(L)\frac{dL}{dD} \propto D^{2+3\alpha}dD.
\label{alpha_beta}
\end{equation}
Therefore the luminosity and size distribution slopes should be related as $\beta=2+3\alpha$. 

The edges which H{\small II}phot uses to determine the total flux of each source are based on the background emission.  For individual spherical \hh regions the FWHM$_{\rm eff}$ is close to half of the assigned border diameter, but in crowded regions especially with  strong background DIG, the assigned borders  are much larger than the FWHM$_{\rm eff}$. The fact that in some regions the reported flux has been integrated on a larger area than the defined effective size may affect the relation between LF and size distribution.
We fit a power law size distribution:
\begin{equation}
	N(D) \propto D^{\beta} dD
\end{equation}
where $D$ is the effective  FWHM size of the region multiplied by two.  

\begin{figure}
\hspace{-1cm}
\center
\includegraphics[width=12cm]{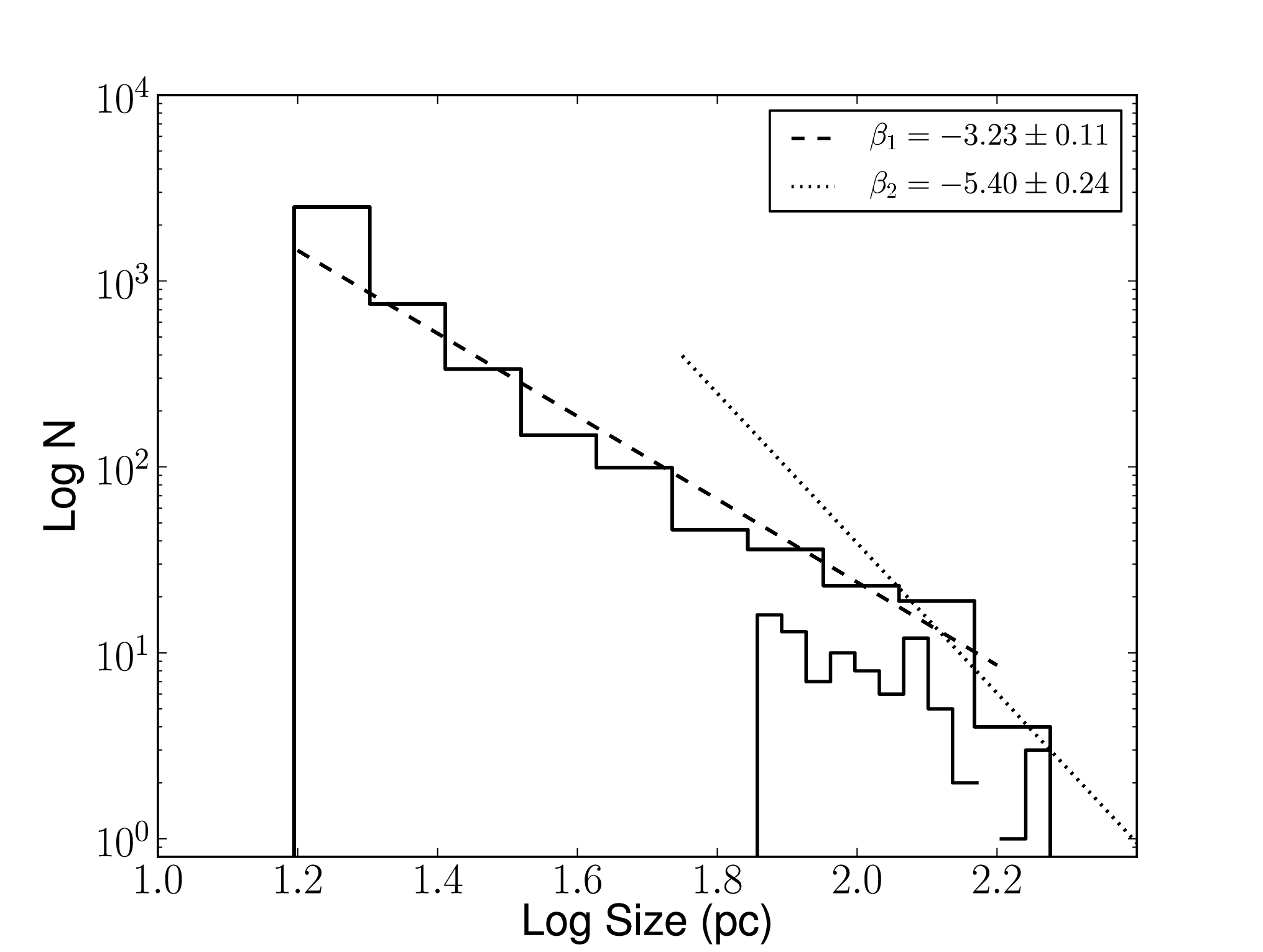}
\caption{
Size distribution of the \hh regions.  
The right corner histogram shows only larger regions ($\log(D{\rm [pc]}) > 1.84$) with 1/3 smaller bins.
The dashed line presents the best power law fit for $D<130$~pc and  dotted line presents the larger end ($D\geq130$~pc) [note that the fitted slope for $\log N (\log D)$ vs. $\log (D)$ is $1-\beta$]. 
\label{SD}}
\end{figure}

Figure \ref{SD} shows the \hh regions' diameter distribution. Similar to \citet{gutier11} we also see a sharp fall for regions larger than 130 pc (the transition between single OB star regions and blends of multiple regions).   The dashed line presents the best fit for all the regions smaller than 130 pc and the dotted line presents the fit for $D>130$~pc. 
We derived a slope of $-3.3\pm0.11$ for the  distribution of the smaller regions and $-5.40\pm0.24$ for the larger end ($D> 130$~pc) of the size distribution.  Due to the galaxy's proximity, most of the larger regions in M31 are resolved, therefore we consider only the large end of the size distribution  for comparison with other galaxies.   Values between $-3.33$ and $-5.50$ have been observed for regions with $\log(D/{\rm pc}) \geq 2.3$ for a range of  Hubble types from 
Sb to Sm-IV \citep[Figure 2 in][]{oey03}, so M31 is consistent with the general trend.
Using Eq.~\ref{alpha_beta} and our value for $\alpha=-2.4\pm 0.17$, we predict $\beta=-5.2\pm0.51$ which is consistent with our fit to
the size distribution. The agreement seems surprisingly good, given that the size that  H{\small II}phot uses to measure the total flux of a region 
is not the same as the effective size (FWHM of the luminosity peak); however, if most of the luminosity of a region is in the central peak rather than the faint outskirts then the agreement is understandable.

\subsection{Comparison of \hh  Regions to Young Clusters in M31} \label{clusters}

It is well accepted that the star formation rate is correlated with the amount of gas in star forming regions  and the gas surface density \citep[e.g.][]{kennicutt98}. Most of the molecular gas lies in spiral arms in a galaxy, therefore it is expected that  the major fraction of the star formation takes place in spiral arms. Most of the stars, and particularly massive stars,   form within clusters and create   \hh regions by ionizing the surrounding gas.  In this section we study the correlation between \hh regions and stellar clusters in M31. We used the Caldwell catalog \citep{caldwell2009} and the  HST/WFPC survey of bright young clusters in M31 \citep{KH07, HK09,HK10,perina2010} to examine the spatial  correlation of \hh regions with young stellar clusters. 

The \citet{caldwell2009} catalog contains 670  clusters of different types, 140 of which are identified as young clusters. 
The HST survey (referred to below as the `KH catalog') and its follow-ups contains 714 clusters, including 67 previously known objects and 24 
clusters in common with Caldwell. 
Only 7 young clusters (age$<$2~Gyr) from the Caldwell catalog were matched within a $10''$ ($\sim38$ pc) neighbourhood (typical size of the Caldwell clusters). The age of the  Caldwell clusters ($\sim $2 Gyr) is much larger than the few Myr lifetime of the \hh regions  so the lack of matches was not a surprise. In contrast, we found an additional 43 \hh regions matched with the KH list within a $4''$ neighbourhood (the typical size of the new KH clusters and the minimum size of our \hh regions). \citet{KH07} used only 39 HST pointings  which cover a very small fraction of the entire disk.  By extrapolating the data they estimated that the entire disk may contain $\approx$80000 stellar clusters. Following their estimation we expect 4370 matches between \hh regions and clusters. This number is larger than our total number of \hh regions; however we already have regions within the KH fields which are not matched with clusters. Overall, we find a reasonable correlation between the location of \hh regions and clusters.

\begin{figure}
\hspace{-1cm}
\center
\includegraphics[width = 15cm]{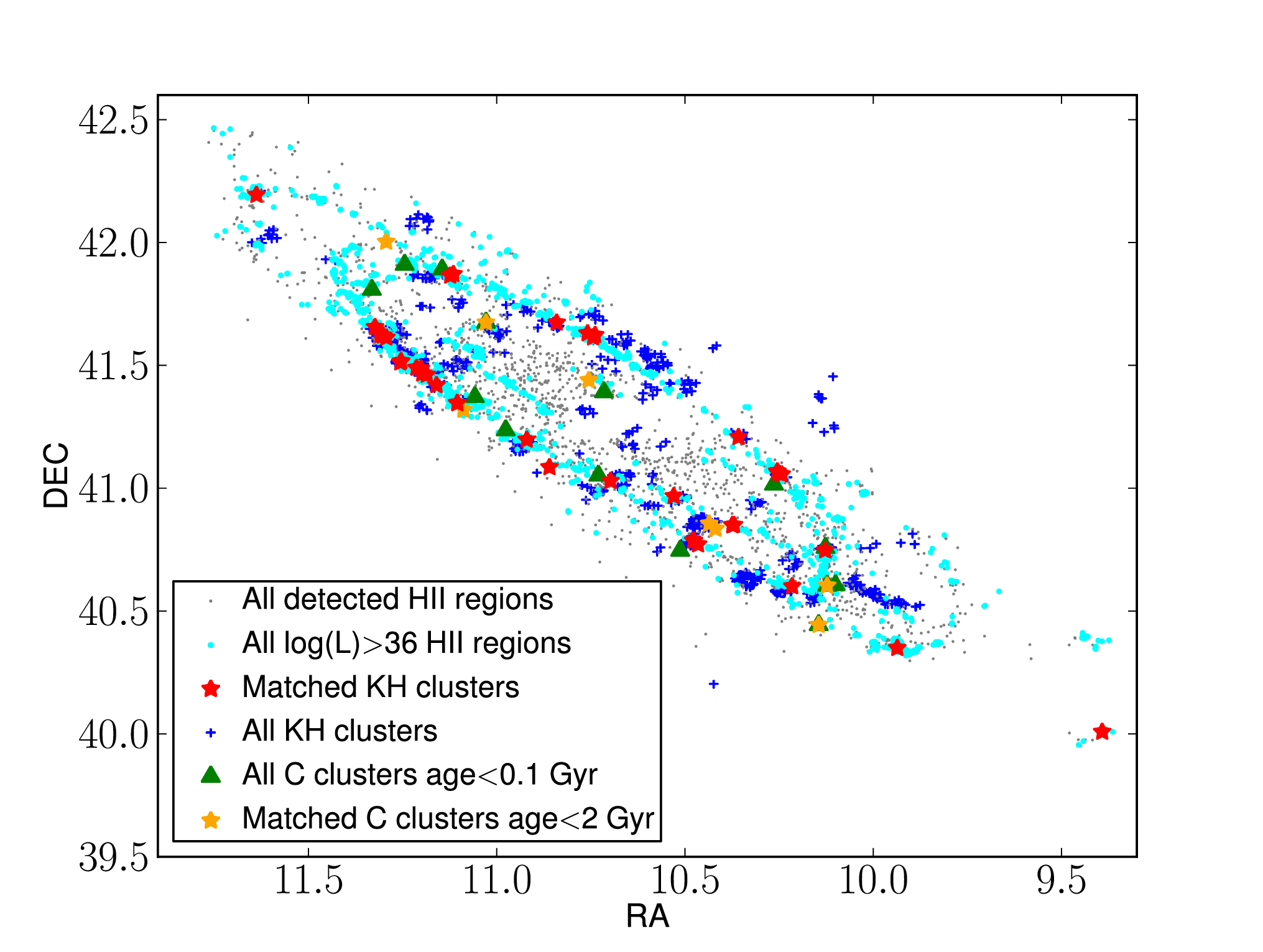}
\caption{  An overlay  of  \hh regions and star clusters in M31. The \hh regions with L$_{H\alpha} \geq 10^{36}$ erg s$^{-1}$ are shown in cyan and other regions in grey dots. The blue crosses represent all the KH clusters.  Green triangles present all the Caldwell young clusters (age $< 0.1$ Gyr). Red and orange stars present the matched \hh regions with the KH and Caldwell (age $<2$ Gyr only) clusters respectively. As expected most of the matches lie in spiral arms. 
 \label{clusters2}}
\end{figure}

Figure \ref{clusters2} compares the locations of the detected \hh regions and matched clusters. Grey and cyan dots present the location of faint and luminous \hh regions. Interestingly, the clusters matched with \hh regions also lie within the spiral arms, indicating  that the clusters within the arms still contain multiple massive stars which can create very luminous \hh regions. 
As discussed above, the luminous \hh regions trace the spiral structure and 10~kpc  star forming ring. The fainter regions are scattered within the arms as well and are more concentrated toward the center (also shown in Figure \ref{regions1}). 
We removed all the known and potential PNe from our catalog, but some of these faint central regions, especially the compact ones, might be unidentified PNe.  The expected lack of active star formation in this gas-poor part of the galaxy is a further argument for some of these objects being PNe associated
with the old stellar population of the bulge.

\citet{schruba2010} recently showed that the Kennicutt-Schmidt law \citep[power law dependence of star formation rate on molecular gas surface density;][]{kennicutt98} which is observed and well accepted for kpc scales breaks down on smaller scales (e.g. aperture size of $\lesssim 300$~pc for M33).  They concluded that individual GMCs present a range of evolutionary states in their $20-30$~Myr lifetime which could be studied by recent (H$\alpha$) and future (CO) star formation tracers. If we extend these results into our observations of recent star formation as detected in \hh regions in highly extincted regions, we may conclude that some very young clusters are missing in this study. 
 Presumably such clusters are deeply embedded in molecular clouds and would be more easily detected at infrared wavelengths, such as with the {\it Herschel Space Observatory} or James Webb Space Telescope.

\section{Summary and Conclusion} 
\label{summary}

We used the data from the Survey of Local Group Galaxies \citep{massey06} to construct a complete catalog of \hh regions in M31. The results of this work are summarized  below:

\begin{enumerate}

\item{Using H$\alpha$ and R band images of M31, we applied the H{\small II}phot code in order to catalog the \hh regions of the galaxy.  The catalog contains 3961  regions to a limiting flux  of    $\sim 10^{-16}$ erg cm$^{-2}$ s$^{-1}$  after removing matched planetary nebulae and potential PNe.  
The luminosity derived for each region was corrected for extinction using {\em Spitzer} MIPS dust maps.
Detected regions with  L$_{H\alpha} \geq 10^{36}$ erg s$^{-1}$ clearly trace the spiral structure of the galaxy while fainter regions scatter through the disk with an increasing number density toward the center. The most luminous region in our catalog, with L$_{H\alpha}=2.2\times10^{38}$ erg s$^{-1}$, confirms that M31 does not have giant \hh regions such as those in the Milky Way, LMC, M33 and M51. }

\item{We measured a total H$\alpha$ luminosity of $5.6\times10^{40}$~erg~s$^{-1}$  which contains a 65\%  contribution from diffuse ionized gas and has about 20\% uncertainty. 
Despite the general belief that small individual   \hh regions only supply a small fraction of the total  H$\alpha$ emission in most  galaxies, 60\% of the total L$_{H\alpha}$ from \hh regions  in M31 is contributed by regions with L$_{H\alpha} < 10^{37}$ erg~s$^{-1}$. This large fraction also contains  the  regions at peripheries of larger complexes that have not been resolved in other studies. In total we determined a SFR=0.44 M$_\odot$~yr$^{-1}$, in agreement with \citet{barmby06} (0.4 M$_\odot$~yr$^{-1}$). }

\item{The luminosity function ($N(L) dL= AL^{\alpha} dL $) and size distribution ($N(D) dD= AD^{\beta} dD$) obtained from the catalog are reasonably fitted by power law distributions.  A slope of $\alpha = -2.52\pm 0.07$ was obtained  for the LF,  which   is consistent with  $-2.3 \pm 0.2$ determined by  \citet{kennicutt89}. 
We detect a break in the size distribution function at $\sim130$~pc, close to 120~pc which is suggested to be the transition between \hh regions ionized by a single OB cluster to multiple regions in M51 \citep{gutier11}. The power law fit to the size distribution for diameters smaller than 130 pc yields a slope of $-3.23\pm0.11$ and a slope of $-5.40\pm0.24$ for $D > 130$~pc,  within the range of the typical distribution for  spiral galaxies but steeper than most Sb galaxies \citep{oey03}. }

\item{The luminosity function has two distinct peaks at approximate L$_{H\alpha}=10^{35}$ and $4\times10^{36}$ erg~s$^{-1}$.  The peak at fainter luminosities is from the population of inter-arm regions and might be contaminated by unidentified PNe.  The second peak luminosity, which also matches with B stars, suggests a starburst 15--20~Myr ago. This timescale is consistent with the results of UV studies which suggest a starburst sometime between 10 and 100~Myr ago. }

\item{Massive stars form in stellar clusters, therefore we examined the correlation between the \hh regions and star clusters. The clusters which spatially match with \hh regions approximately lie within the arms. Considering  the incompleteness of the M31 young cluster catalogs, we found a good statistical match between the location of \hh regions and clusters. }

\end{enumerate}

\acknowledgments

We thank Fatemeh Tabatabaei for providing the extinction map for M31 and Johan Knapen for useful comments and suggestions
on a draft of the paper.  Nelson Caldwell and Philip Massey are thanked for helpful discussions and suggestions. 
We also thank the anonymous referee for detailed comments and suggestions that helped to improve this work.
Support for this research was provided by a Discovery Grant from the Natural Sciences
and Engineering Research Council of Canada and an Ontario Early  Researcher Award, both to PB.

\newpage

\newpage
\begin{deluxetable}{cc}
\tabletypesize{\scriptsize}
\tablecaption{Scaling factors for H$\alpha$ continuum subtraction of Local Group Survey M31 images\label{scale}}
\tablewidth{0pt}
\tablehead{
\colhead{Field}& \colhead{Scaling factor}   }
\startdata
1 & 0.378 \\
2 & 0.855\\
3 & 0.379\\
4 & 0.395\\
5 & 0.363\\
6 & 0.393\\
7 & 0.386\\
8 & 0.418\\
9 & 0.400\\
10 & 0.396\\
\enddata
\tablecomments{Scaling factors are the average ratio of H$\alpha$ to $R$-band total flux (as measured in
the original image units) of bright stars in each field. See text for discussion of field 2.}
\end{deluxetable}

\begin{deluxetable}{lccccccccccccc}
\rotate
\tabletypesize{\scriptsize}
\tablecaption{Catalog of \hh regions in M31 \label{tb1} }
\tablewidth{0pt}
\tablehead{
\colhead{ID}&\colhead{Field}&\colhead{RA$_{(J2000)}$}&\colhead{DEC$_{(J2000)}$} & \colhead{FWHM$_{maj} $}& \colhead{FWHM$_{min}$ }& \colhead{PA} 
&\colhead{D} &\colhead{Flux ($\times10^{-15}$)} & \colhead{Extinction} &\colhead{Luminosity$_{cor}$} \\
\colhead{}&\colhead{}&\colhead{(deg)}&\colhead{(deg)}&\colhead{(\arcsec)}&\colhead{(\arcsec)}&\colhead{(deg)}&\colhead{(pc)}&\colhead{(erg~cm$^{-2}$~s$^{-1}$)}&\colhead{(mag)}&\colhead{$\times10^{34}$(~erg~s$^{-1}$)}}
\startdata

904 & F9 & 10.2175 & 40.5493 & 4.65 & 2.66 & 105 & 26.67 & 19.50$\pm$1.67 & 1.341 & 492\tabularnewline
905 & F7 & 10.2175 & 40.6027 & 3.64 & 2.08 & 135 & 20.85 & 7.32$\pm$1.93 & 0.892 & 122\tabularnewline
906 & F7 & 10.2175 & 40.6224 & 2.78 & 1.85 & 75 & 17.23 & 4.63$\pm$2.38 & 2.396 & 309\tabularnewline
907& F9 & 10.2179 & 40.5306 & 2.79 & 2.23 & 90 & 18.95 & 1.78$\pm$0.71 & 0.664 & 23.9\tabularnewline
908 & F8 & 10.2183 & 40.548 & 18.04 & 14.43 & 120 & 122.43 & 68.20$\pm$1.69 & 1.341 & 1720\tabularnewline
909 & F8 & 10.2183 & 40.6016 & 3.21 & 1.61 & 0 & 17.23 & 16.50$\pm$0.19 & 0.771 & 246\tabularnewline
910 & F8 & 10.2187 & 40.5547 & 2.54 & 2.03 & 90 & 17.23 & 0.76$\pm$0.36 & 1.265 & 17.8\tabularnewline
911 & F7& 10.2196 & 41.0024 & 4.66 & 4.66 & 0 & 35.40 & 3.17$\pm$0.90 & 1.072 & 62.3\tabularnewline
912 & F8 & 10.2204 & 40.5397 & 4.27 & 2.14 & 60 & 22.93 & 19.90$\pm$0.91 & 1.018 & 374\tabularnewline
913 & F9 & 10.2204 & 40.5473 & 3.53 & 1.77 & 165 & 18.95 & 6.39$\pm$0.99 & 1.227 & 145\tabularnewline
914 & F8& 10.2208 & 40.5383 & 2.79 & 2.23 & 45 & 18.95 & 60.30$\pm$1.04 & 0.914 & 103\tabularnewline
915 & F8& 10.2208 & 40.7258 & 2.27 & 2.27 & 0 & 17.23 & 4.20$\pm$0.86 & 1.239 & 96.4\tabularnewline
916 & F7 & 10.2212 & 40.9862 & 3.89 & 1.94 & 45 & 20.85 & 9.03$\pm$1.38 & 0.952 & 159\tabularnewline
917 & F7 & 10.2212 & 41.0138 & 2.31 & 1.85 & 0 & 15.66 & 0.41$\pm$0.33 & 1.499 & 12.0\tabularnewline
918 & F9 & 10.2217 & 40.5484 & 5.45 & 2.73 & 165 & 29.25 & 3.65$\pm$0.85 & 1.227 & 82.8\tabularnewline
919 & F8 & 10.2229 & 40.5375 & 2.78 & 1.85 & 60 & 17.23 & 1.75$\pm$0.34 & 0.914 & 29.9\tabularnewline
920 & F8& 10.2246 & 40.6179 & 4.84 & 2.76 & 135 & 27.74 & 13.30$\pm$0.22 & 2.409 & 899\tabularnewline
921 & F8& 10.2263 & 40.6409 & 2.27 & 2.27 & 0 & 17.23 & 0.52$\pm$0.04 & 1.244 & 11.9\tabularnewline
922 & F7 & 10.2263 & 40.7661 & 5.85 & 3.34 & 105 & 33.58 & 10.90$\pm$2.90 & 1.149 & 231\tabularnewline
923 & F8 &10.2271&	40.6194 & 2.54 & 2.03 & 90 & 17.23 & 6.46$\pm$0.17 & 2.637 & 537\tabularnewline
\enddata
\end{deluxetable}

\begin{deluxetable}{lccccccccccccc}
\rotate
\tabletypesize{\scriptsize}
\tablecaption{Catalog of potential PNe, removed from final \hh region  catalog \label{removed} }
\tablewidth{0pt}
\tablehead{
\colhead{ID}&\colhead{Field}&\colhead{RA$_{(J2000)}$}&\colhead{DEC$_{(J2000)}$} & \colhead{FWHM$_{maj} $}& \colhead{FWHM$_{min}$ }& \colhead{PA} 
&\colhead{D} &\colhead{Flux ($\times10^{-15}$)} & \colhead{Extinction} &\colhead{Luminosity$_{cor} $} &\colhead{\citet{merrett06}}\\
\colhead{}&\colhead{}&\colhead{(deg)}&\colhead{(deg)}&\colhead{(\arcsec)}&\colhead{(\arcsec)}&\colhead{(deg)}&\colhead{(pc)}&\colhead{(erg~cm$^{-2}$~s$^{-1}$)}&\colhead{(mag)}&\colhead{$\times10^{34}$(~erg~s$^{-1}$)}& \colhead{Catalog number}}
\startdata
100 & F8 & 10.2196 & 40.6769 & 2.538 & 2.031 & 45 & 17.230 & 0.671$\pm$0.0387 & 0.503 & 7.81 & 2918\tabularnewline
101 & F7 & 10.2242 & 41.023 & 2.308 & 1.846 & 0 & 15.664 & 0.909$\pm$0.344 & 1.368 & 23.5 & 1550\tabularnewline
102 & F8 & 10.2263 & 40.5382 & 2.308 & 1.846 & 0 & 15.664 & 0.743$\pm$0.259 & 0.839 & 11.8 & 2264\tabularnewline
103 & F6 & 10.2263 & 41.1441 & 2.064 & 2.064 & 0 & 15.664 & 2.1$\pm$0.161 & 0.000 & 15.4 & 3041\tabularnewline
104 & F8 & 10.2317 & 40.7221 & 2.538 & 2.031 & 135 & 17.230 & 0.549$\pm$0.24 & 1.043 & 10.5 & 1907\tabularnewline
105 & F7 & 10.235 & 40.8982 & 2.064 & 2.064 & 0 & 15.664 & 2.49$\pm$0.169 & 0.390 & 26.2 & 1778\tabularnewline
106 & F7 & 10.2358 & 40.7829 & 2.064 & 2.064 & 0 & 15.664 & 3.55$\pm$0.904 & 1.048 & 68.4 & 1923\tabularnewline
107 & F7 & 10.245 & 40.8629 & 2.308 & 1.846 & 0 & 15.664 & 0.977$\pm$0.238 & 0.632 & 12.8 & 1772\tabularnewline
108 & F7 & 10.2471 & 41.0494 & 2.308 & 1.846 & 0 & 15.664 & 2.95$\pm$0.486 & 1.438 & 81.5 & 1556\tabularnewline
109 & F8 & 10.2513 & 40.4805 & 2.064 & 2.064 & 0 & 15.664 & 1.86$\pm$0.256 & 0.000 & 13.6 & 2256\tabularnewline
110 & F7 & 10.2513 & 40.8982 & 2.918 & 1.460 & 105 & 15.664 & 0.543$\pm$0.149 & 0.429 & 5.91 & 2877\tabularnewline
111 & F7 & 10.2613 & 41.1157 & 2.064 & 2.064 & 0 & 15.664 & 1.22$\pm$0.347 & 0.874 & 19.9 & 1570\tabularnewline
112 & F8 & 10.2742 & 40.4704 & 2.538 & 2.031 & 120 & 17.230 & 0.568$\pm$0.188 & 0.000 & 4.16 & 2251\tabularnewline
113 & F7 & 10.2808 & 40.843 & 2.064 & 2.064 & 0 & 15.664 & 1.98$\pm$1.21 & 0.627 & 25.9 & 2876\tabularnewline
114 & F6 & 10.3058 & 41.193 & 2.064 & 2.064 & 0 & 15.664 & 0.722$\pm$0.153 & 0.847 & 11.5 & 1033\tabularnewline
115 & F7 & 10.3083 & 40.7816 & 2.064 & 2.064 & 0 & 15.664 & 2.76$\pm$1.41 & 0.594 & 35 & 1922\tabularnewline
116 & F7 & 10.3112 & 40.7425 & 2.064 & 2.064 & 0 & 15.664 & 5.63$\pm$3.22 & 0.000 & 41.3 & 1910\tabularnewline
117 & F6 & 10.3217 & 41.2672 & 2.064 & 2.064 & 0 & 15.664 & 1.22$\pm$0.104 & 0.546 & 14.7 & 1053\tabularnewline
118 & F6 & 10.3229 & 41.2018 & 2.064 & 2.064 & 0 & 15.664 & 1.56$\pm$0.158 & 0.969 & 27.9 & 1039\tabularnewline
119 & F7 & 10.3233 & 40.9791 & 2.064 & 2.064 & 0 & 15.664 & 2.06$\pm$0.592 & 0.546 & 24.9 & 1760\tabularnewline
120 & F6 & 10.3363 & 41.1714 & 2.064 & 2.064 & 0 & 15.664 & 1.76$\pm$ 0.11 & 1.382 & 45.9 & 1029\tabularnewline

\enddata
\end{deluxetable}

\newpage

\begin{deluxetable}{lllc}
\tabletypesize{\scriptsize}
\tablecaption{Star formation rates for M31\label{SFR}}
\tablewidth{0pt}
\tablehead{
\colhead{Data}& \colhead{Method} & \colhead{SFR (M$_\odot$ yr$^{-1}$)}  
}
\startdata
\cite{Taba2010}  & extinction corrected H$_\alpha$& 0.27--0.38\\
\citet{kang09} & UV SF regions, 400 Myr avg & 0.6--0.7 \\
\citet{barmby06} & Infrared  8$\mu$m Luminosity&0.4\\
\citet{williams03_1} &Optical photometry& $\sim1$\\
\citet{walterbos94} & extinction corrected H$_\alpha$ & 0.35\\
\enddata
\end{deluxetable}

\clearpage
\newpage
\bibliographystyle{aas}
\bibliography{bibp3}
\end{document}